\documentclass{article}

\PassOptionsToPackage{numbers,sort&compress}{natbib}

\usepackage[preprint]{neurips_2026}

\usepackage[utf8]{inputenc}
\usepackage[T1]{fontenc}
\usepackage{url}
\usepackage{booktabs}
\usepackage{graphicx}
\usepackage{amsmath}
\usepackage{amssymb}
\usepackage{nicefrac}
\usepackage{microtype}
\usepackage{xcolor}
\usepackage{float}
\usepackage{placeins}
\usepackage{newunicodechar}
\newunicodechar{≈}{\ensuremath{\approx}}

\usepackage{hyperref}
\hypersetup{colorlinks=true, linkcolor=blue, citecolor=blue, urlcolor=blue}

\title{Probing and steering biology across Boltz-1}

\author{%
  Piotr Jedryszek\textsuperscript{1, 2}\thanks{Corresponding author: \texttt{piotrjedryszek@evolverebiosciences.com}. Affiliations: \textsuperscript{1}Dept.\ of Biology, Univ.\ of Oxford; \textsuperscript{2}Evolvere Biosciences, London; \textsuperscript{3}Kavli Institute, Univ.\ of Oxford; \textsuperscript{4}Dept.\ of Chemistry, Univ.\ of Oxford; \textsuperscript{5}Dept.\ of Oncology, Univ.\ of Oxford; \textsuperscript{6}Botnar Research Centre, NDORMS, Univ.\ of Oxford; \textsuperscript{7}Independent Researcher.} \quad
  Tongmeng Xie\textsuperscript{7} \quad
  Adam Winnifrith\textsuperscript{2,6} \quad
  Alexander Hasson\textsuperscript{2,5} \\
  \textbf{Weronika Slesak\textsuperscript{1, 2}} \quad
  \textbf{George Wicks\textsuperscript{2}} \quad
  \textbf{Toby Winnifrith\textsuperscript{2}} \quad
  \textbf{Oliver M. Crook\textsuperscript{3, 4}}
}

\begin{document}
\maketitle

\begin{abstract}
AlphaFold3-class structure predictors pair a representational trunk, which processes sequence and context, with a diffusion module, which generates atomic coordinates. How biological information changes as it crosses this architectural boundary remains poorly understood. We analyze per-residue activations from the Pairformer trunk and diffusion module of Boltz-1 using linear probes, sparse autoencoders (SAEs), and causal interventions. From the trunk, both geometry (secondary structure, disorder) and sequence chemistry (amino-acid identity, signal peptides, disulfide-bond annotations) are linearly decodable. In the diffusion module, the two diverge. Secondary structure remains equally decodable, whereas sequence chemistry becomes much less linearly accessible. We then test whether decodable directions can steer the model by editing the trunk's single representation.  Helix and coil directions change predicted structure dose-dependently, but a $\beta$-strand direction that is highly predictive on the same proteins (held-out F1 $=0.78$) produces no measurable increase in strand content at any of the steered layers. Removing helix or coil directions does not remove that structure. Linear decodability therefore did not imply causal influence at the sites we tested. The same probes also score markedly lower against SwissProt annotations than against DSSP labels, because unannotated residues that the model gets right are charged as false positives; such scores are therefore lower bounds. Finally, supervised probes outscore single SAE features wherever a label already exists. We release the trained trunk and diffusion SAEs,\footnote{\href{https://huggingface.co/collections/evolve-away/boltz-saes}{Hugging Face collection of SAEs}} Boltz-1 per-residue activations, and the analysis code.\footnote{\href{https://github.com/EvolvereLabs/Boltz-interp}{GitHub repository, with instructions for activation download}}
\end{abstract}

\section{Introduction}
\label{sec:intro}

Protein structure predictors such as AlphaFold3 \citep{abramson_accurate_2024} and Boltz-1 \citep{wohlwend_boltz-1_2024} separate representation learning from coordinate generation. Boltz-1's Pairformer trunk iteratively updates a per-residue single representation, $s$, and a residue-pair representation, $z$; a diffusion module then generates atomic coordinates conditioned on the trunk output. This motivates our central question: how do the trunk and diffusion module differ in the biological information they represent?

Related work has looked inside individual modules. \citet{feldman_alphainterp_2026} argue AlphaFold3's Pairformer compresses evolutionary information into an increasingly linearly-decodable latent geometry. \citet{lu_two_2026} combine activation patching, probes, and steering in the folding trunks of ESMFold, OpenFold, and Boltz-1, finding that early blocks write biochemical signals such as charge into the pair representation and later blocks build pairwise geometry; steering charge or pairwise-distance directions induces $\beta$-hairpins in all three models. \citet{parsan_towards_2025} and \citet{zarzecki_foldsae_2026} steer ESMFold and a generative structure model, respectively, with SAE features. Concurrently with our work, \citet{kim_bish-bash-fold_2026} scale SAEs to the diffusion modules of Boltz-2 and AlphaFold3 and likewise find a geometric rather than biochemical focus, consistent with what we report here for Boltz-1's diffusion module. To our knowledge, no prior work has systematically contrasted the trunk/diffusion boundary itself, nor tested whether directions that decode per-residue biological annotations from the trunk output steer the structure the diffusion module generates.

We make four contributions. First, evaluating per-residue activations from both modules against secondary-structure, sequence-chemical, and amino-acid-identity labels, we show the Pairformer trunk keeps both geometry and sequence chemistry linearly accessible, while the diffusion module keeps geometry but not chemistry. Second, we causally test the trunk's decodable secondary-structure signal by editing its single representation at thirteen layers: helix and coil directions steer the predicted structure, but a highly decodable strand direction does not at any depth, and removing a direction does not remove the secondary structure it encodes. Decodability and causal steerability are therefore distinct properties. Third, we show that the common evaluation practice of scoring against sparse curated annotations can inflate the false-positive rate, because missing labels turn correct predictions into apparent errors, making probes, SAE directions, and other interpretability units look less aligned with the labeled concept than they really are. Fourth, on both decoding and steering, supervised probes outperform single SAE features wherever a label already exists. More broadly, our decoding, intervention, and annotation controls identify practical failure modes for attempts to extract biological knowledge from model representations.

\section{Methods}
\label{sec:methods}

\textbf{Model and activations.} We run Boltz-1 with precomputed multiple sequence alignments (MSAs) from the OpenFold set \citep{ahdritz_openfold_2022}. From the Pairformer trunk (48 layers, recycle 1), we extract per-residue activations at every second layer (layers 0--47). From the diffusion module, we extract the token representation (one token per standard residue) at token-transformer layers 0, 2, 4, 6, 10, 14, 18, and 22 and sampling steps 0, 10, 50, 100, and 199 of 200, yielding a layer-by-step grid. Because the two modules differ in depth, cross-module comparisons index activations by fractional depth (layer index / total layers) and are restricted to a matched set of evaluation proteins. Probes and SAEs are always trained separately per module; none are shared across the trunk/diffusion boundary. Full extraction details are in Appendix~\ref{app:methods_sup}.

\textbf{Datasets and labels.} SAEs are trained on 84,074 unlabeled proteins (21.9M residues). Evaluation uses a fixed 486-protein set with three overlapping label sources: \emph{sparse} SwissProt annotations (99 proteins), which cover only experimentally characterized regions and leave most residues unlabeled; \emph{dense} DSSP labels, which assign every residue a helix, strand, or coil state from a 3D structure, computed from AlphaFold DB structures (393 proteins); and dense DSSP labels computed from Boltz-1's own predicted structures (all 486), generated with our activation-extraction settings and used as a self-consistency control. We group concepts into \textbf{geometry} (helix, strand, coil, disorder) and \textbf{sequence chemistry} (amino-acid identity, signal peptide, disulfide bond); amino-acid identity is a positive control. Details are in Table~\ref{tab:dataset_label_prevalence}.

\textbf{Sparse autoencoders and readouts.} We train TopK SAEs ($k=256$, 2048 latents, 3 seeds) with $\ell_2$ weight regularization \citep{jedryszek_stable_2026} on demeaned activations, that is, with each dimension's training-set mean subtracted, because a few very high-variance dimensions otherwise dominate learning. We compare four \textbf{readouts}: (a) \emph{probe-raw}, a logistic probe fit on raw activations and scored on held-out proteins; (b) \emph{probe-SAE}, the same probe applied to SAE latents; (c) \emph{SAE-1feat}, the single best SAE latent for a concept; (d) \emph{neuron-1feat}, the single best raw neuron. All scores are F1, where $1.00$ is perfect, and permutation nulls sit near zero. Full probe/F1/null protocols are in Appendix~\ref{app:methods_sup}.

\textbf{Causal steering.} Steering tests whether a decodable direction is actually used: adding it tests whether it is sufficient to change the prediction, and removing it tests whether the model needs it. We steer secondary structure because DSSP gives a dense, automatic readout of it on the predicted structure with a clear expected direction; concepts such as signal peptides have no such readout. We add a unit direction $u$ (either a supervised probe direction or an SAE decoder direction) to the trunk single representation, $s$, at the trunk output that conditions the diffusion module and, in a depth sweep, at earlier Pairformer layers; the pair representation, $z$, is not modified. We split the dataset into 389 proteins for fitting directions and 97 held-out proteins for testing, to avoid circularity. We report each edit's effect on the held-out proteins as the paired per-protein difference from a control edit of the same size, with bootstrap confidence intervals (CIs); additions are also compared against a null of 19 random directions. Full protocol in Appendix~\ref{app:steering}.

\section{Results}

\subsection{The trunk and diffusion module exhibit a representational split}
\label{sec:r_split}

We find that the trunk represents both geometry and sequence chemistry. \emph{Probe-raw}, a logistic probe fit on raw activations, decodes geometry strongly at every trunk depth: helix, strand, and coil reach F1 scores of $0.79$--$0.90$, and disorder $0.86$. Sequence chemistry is also accessible, though less strongly, with signal peptides at $0.76$ and disulfide-bond residues at $0.43$ (Appendix Fig.~\ref{fig:trunk}). Amino-acid identity, our positive control, reaches $0.996$ at the trunk output, confirming that labels and activations are aligned (Appendix Figs.~\ref{fig:calibration}, \ref{fig:s1_peraa}).

\begin{figure}[!h]\centering
  \includegraphics[width=\textwidth]{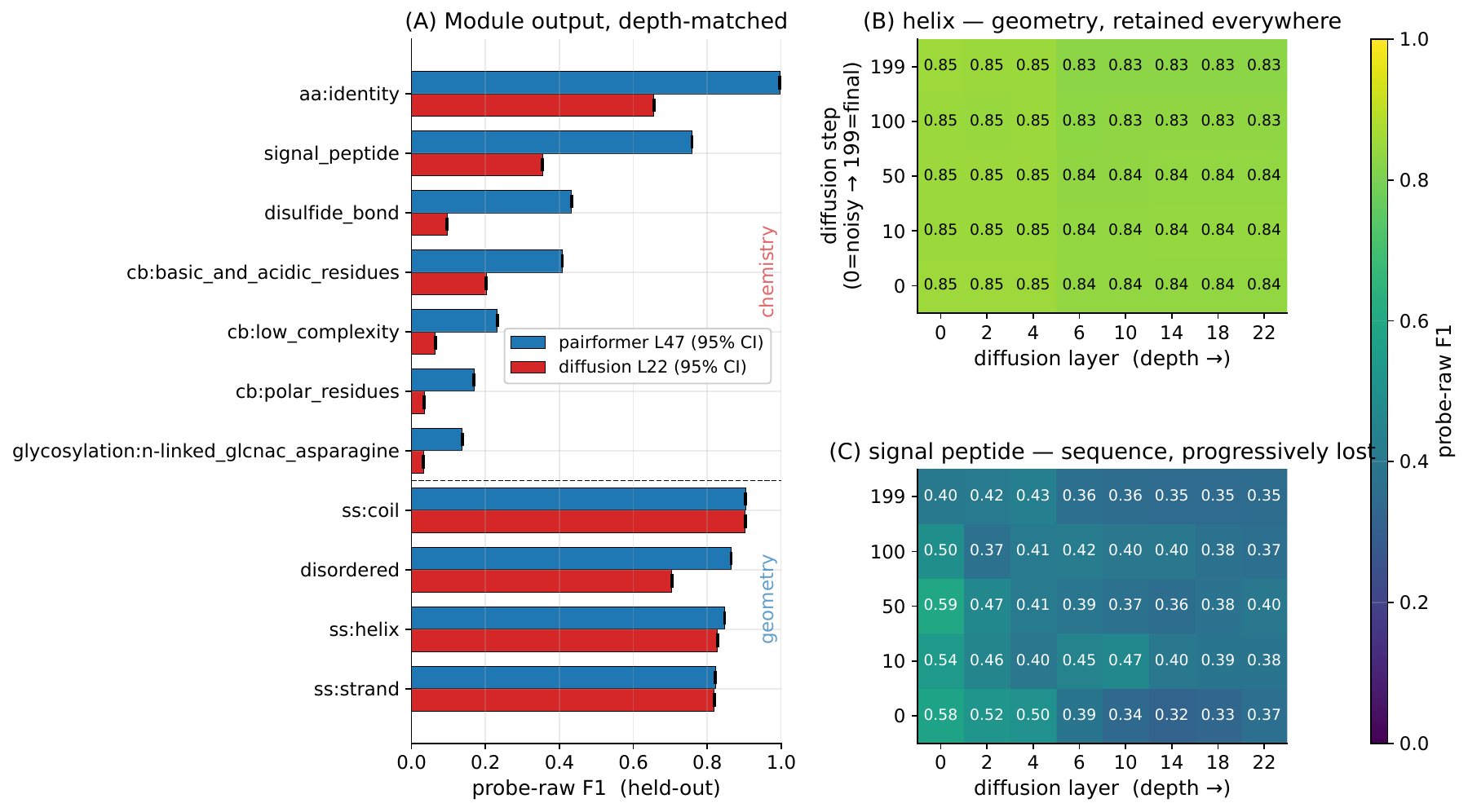}
  \caption{\textbf{The diffusion module keeps geometry linearly accessible while chemistry fades.}
  (A) For depth-matched module outputs (Pairformer L47 vs.\ diffusion L22), geometric features are decodable from both the Pairformer trunk and the diffusion module. In contrast, sequence-chemistry concepts are more decodable in the trunk. (B,C) Diffusion probe-raw over depth
  $\times$ sampling step: helix (B) is bright across the whole grid, already resolved at the
  noisiest step; signal peptide (C) is weakly decodable and fades along both the layer and
  sampling-step axes.}
  \label{fig:split}
\end{figure}

By contrast, the diffusion module's token representation keeps geometry linearly accessible but not sequence chemistry. Comparing each module's deepest extracted layer (trunk L47 vs.\ diffusion L22), secondary structure is essentially preserved, with helix, strand, and coil shifting by at most $0.02$, whereas signal peptide falls from $0.76$ to $0.35$, disulfide bond from $0.43$ to $0.10$, and amino-acid identity from a perfect $1.00$ to $0.66$ (Fig.~\ref{fig:split}A, Table~\ref{tab:headline}). Although the signal-peptide and disulfide labels cover only 99 proteins, both drops exceed sampling uncertainty, with non-overlapping 95\% bootstrap CIs between the two modules (Appendix Fig.~\ref{fig:s11_probe_CI}). The same split holds across the diffusion module's whole depth$\times$step grid, where helix remains decodable even at the noisiest step while signal peptide fades along both axes (Fig.~\ref{fig:split}B,C).

\subsection{Some but not all features are causally steerable in the single representation}
\label{sec:r_steer}

\begin{figure}[t]\centering
  \begin{minipage}[c]{0.46\textwidth}\centering
    \includegraphics[width=\linewidth]{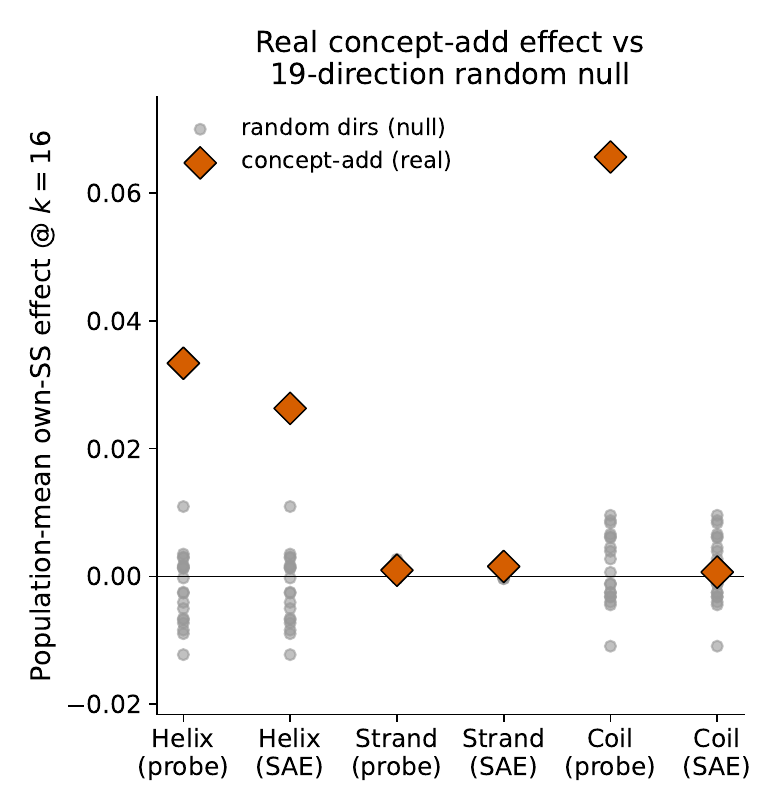}
  \end{minipage}\hfill
  \begin{minipage}[c]{0.54\textwidth}\centering
    \includegraphics[width=\linewidth]{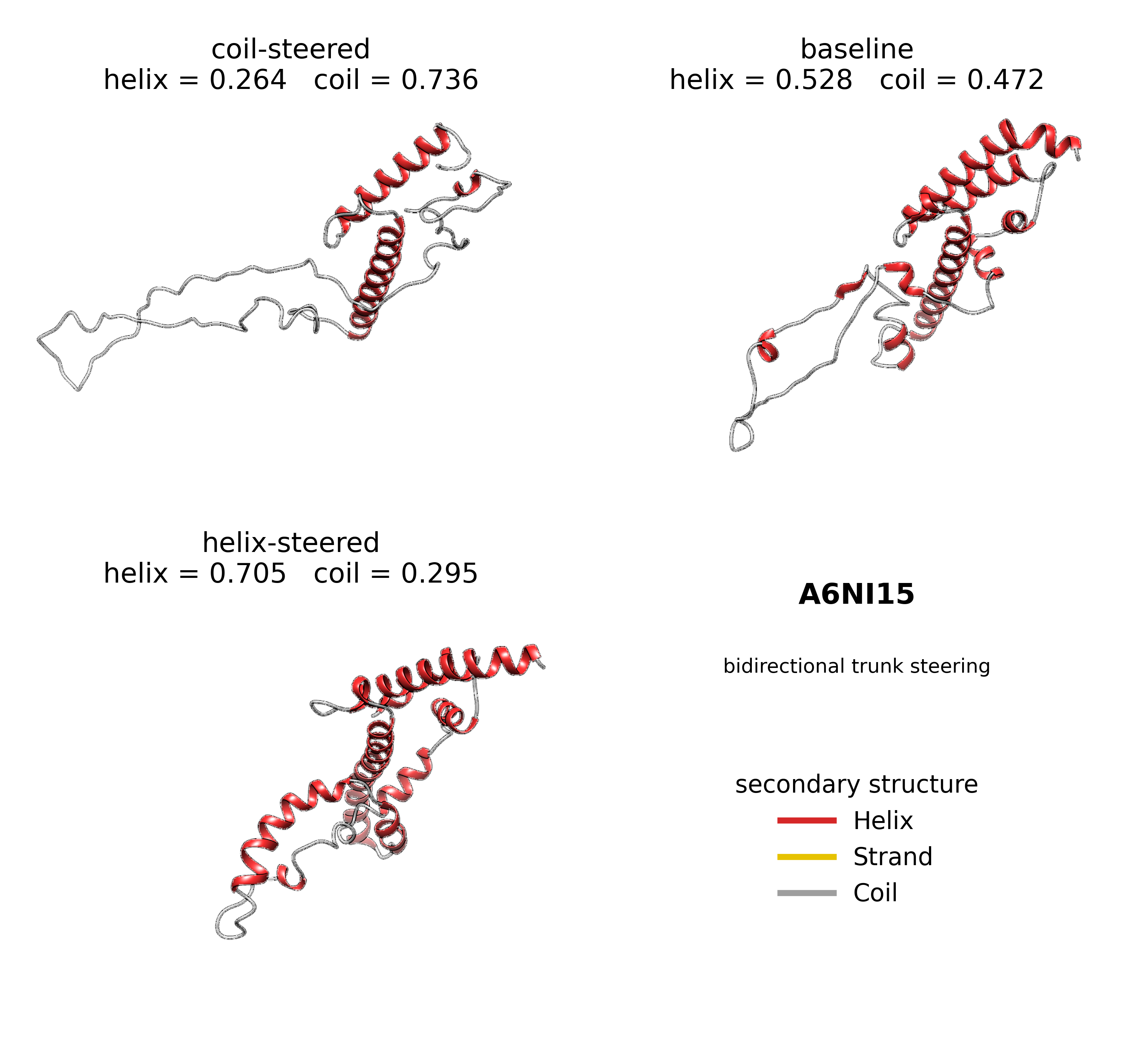}
  \end{minipage}
  \caption{\textbf{Which trunk directions causally steer, and one protein driven both ways.}
  Per-concept population own-state effect ($k{=}16$, $n{=}97$) against a 19-direction random null;
  and held-out A6NI15 under baseline, helix, and coil steering.}
  \label{fig:steer_null}
\end{figure}

We find that adding a helix or coil direction to the trunk output increases that secondary structure in the predicted structure. This was observed for both supervised \textit{probe-raw} directions and SAE decoder direction.  We fit each direction on the DSSP labels of 389 training proteins and added it to $s$ for the 97 held-out proteins. The dose $k$ sets the size of the added vector as a multiple of how far a typical residue's activation sits from the average along that direction, so $k{=}1$ adds one such typical deviation. At $k{=}16$, the helix probe direction raises helix fraction (the share of residues DSSP assigns to helix) by $0.039$ relative to a matched-norm random direction, and the coil probe direction raises coil fraction by $0.061$ (Table~\ref{tab:steering}, Fig.~\ref{fig:steer_null}). Both effects grow with dose while mean pLDDT, the model's per-residue confidence on a 0--100 scale, stays at $\sim$74 (Appendix Fig.~\ref{fig:s_dose}). The shift is modest across the population but several-fold larger in low-confidence structures, suggesting that the conditioning matters most where the fold is not yet settled (Appendix Fig.~\ref{fig:steer_confidence}). Driving one held-out protein each way swaps its helix and coil content (Fig.~\ref{fig:steer_null}, right; Appendix Fig.~\ref{fig:steer_confusion}).

We next asked whether helix and coil steering depended on where the edit was applied. We fitted probe directions at 13 Pairformer layers and tested each at its own layer on the same 97 held-out proteins, at $k{=}16$ and $k{=}64$. Because the same $k$ produces much smaller edits relative to the activations in early layers, we compared layers using \emph{relative edit size}: the norm of the added vector divided by the mean norm of the residue activations.

At $k{=}16$, coil first showed a detectable response at layer 24. Editing this layer increased coil fraction by $0.094$, compared with $0.061$ when editing the last Pairformer layer (layer 47), despite similar relative edit sizes. For helix, no earlier layer produced significantly stronger steering than the last layer after correction for multiple testing (Appendix Fig.~\ref{fig:s_depth}).

Layer 0, however, showed little response even when the relative edit size was comparable to that used successfully at the last layer: helix fraction changed by just $+0.002$, compared with $+0.039$ when editing the last layer. A further dose sweep on 16 held-out proteins again found no detectable layer-0 response, while layer 12 responded at higher doses (Appendix Fig.~\ref{fig:s_depth_dose}). This follow-up used a shorter, more helix-rich subset of the evaluation proteins. The layer-0 null is consistent with subsequent trunk processing suppressing the perturbation, although these experiments do not establish where or how this occurs.

In contrast, a strand probe direction does not additively steer at any of these depths, even though it decodes strand well. Re-scored on the 97 proteins it is added to, this direction identifies strand residues with held-out F1 $0.78$ at precision $0.93$. Added at the trunk output, it leaves strand fraction unchanged ($-0.001$) and instead converts helix into coil. Across all 13 layers it never raises strand fraction by more than $0.003$ at $k{=}16$, over ten times less than the helix and coil effects at the output (Appendix Fig.~\ref{fig:s_depth}). We hypothesize that the architecture explains this null. $\beta$-strand pairing is a property of residue pairs, carried in the pair representation $z$, and inside Boltz-1's Pairformer information flows only from $z$ to $s$, so an edit to $s$ cannot reach it. This is consistent with \citet{lu_two_2026}, who induce $\beta$-hairpins in Boltz-1 by steering pairwise-distance directions in $z$. Their intervention targets contacts between specific residue pairs, whereas our edits act only on $s$ and cannot update $z$ within the Pairformer. Linear decodability therefore does not imply causal influence at the sites we tested.

Finally, we checked whether removing a helix or coil direction from the trunk output reduces the fraction of residues in that state, and found that it does not. We first removed each probe direction from $s$ by setting every residue's component along it to its average value, so the direction no longer differs between residues. Relative to removing a random direction in the same way, this changes helix and coil fraction by at most $0.004$, an order of magnitude below the effect of adding them. However, removing one direction leaves every other direction in $s$ untouched, and the state may still be encoded along those. A stronger intervention of equalising the helix, strand and coil class means in $s$ also left overall helix and coil content largely unchanged. However, a separate diagnostic found that secondary-structure labels remained recoverable by quadratic readouts after analogous erasure, limiting what this null result establishes about the necessity of secondary-structure information (Appendix~\ref{app:necessity}).

\subsection{Sparse curated annotations underestimate what the model represents}
\label{sec:r_annot}

We find that sparse curated annotations make the trunk look far less informative than dense labels do. Across readouts, recall usually exceeds precision (Appendix Fig.~\ref{fig:pr}; DSSP strand is the consistent exception), so we asked whether the low precision is a property of the representation or of the labels. On identical trunk activations, at the layer where each label source decodes best, helix decodes at F1 $0.85$ against dense DSSP labels but $0.41$ against SwissProt's sparse experimental annotations, and strand at $0.83$ against $0.33$ (Appendix Fig.~\ref{fig:underannotation}). Most of these apparent errors are correct predictions: $90\%$ (95\% CI $82$--$95\%$) of residues counted as helix false positives against SwissProt are helices in DSSP of the AlphaFold DB structures, against a $36\%$ base rate among unannotated residues (strand: $61\%$, CI $55$--$66\%$, vs.\ $10\%$; all 99 proteins). Computing DSSP on Boltz-1's own predicted structure instead of the AlphaFold structure changes helix probe-raw by only $\sim$0.06--0.08 (Appendix Fig.~\ref{fig:s7_selfconsistency}), so the gap reflects missing SwissProt labels rather than structural disagreement between the two models. Scores against sparse annotations are therefore lower bounds on what the model represents.

\subsection{Supervised probes outperform single SAE features where labels exist}
\label{sec:r_sae}

We find that, when labels exist, individual SAE latents are weaker readouts and steering directions than a supervised probe. On the trunk's secondary-structure concepts, \emph{probe-raw} beats \emph{SAE-1feat}, the best single SAE latent, by $0.12$--$0.32$ F1 at the trunk output (helix $0.85$ vs.\ $0.73$; strand $0.82$ vs.\ $0.51$), and probes steer more reliably: the coil probe moves coil content while the best coil latent does not (Table~\ref{tab:steering}; Figs.~\ref{fig:pr}, \ref{fig:s5_faithfulness}). We compare against single latents because that is how SAE features are usually read and steered. The concepts are still present in the SAE as a whole: \emph{probe-SAE}, a probe trained on all SAE latents together, recovers $\sim$97\% of probe-raw F1 in the trunk ($\sim$82\% in diffusion). SAE latents may still hold value by surfacing candidate features that existing annotations lack. In a companion study, auto-interpretability on these trunk SAEs recovers functional latents absent from SwissProt, such as a kinase catalytic-histidine motif \citep{jedryszek_efficient_2026}.

\section{Discussion}
\label{sec:discussion}

The representational split (\S\ref{sec:r_split}) reveals a difference in how Boltz-1's Pairformer trunk and diffusion module represent biological information. Why would representations of sequence chemistry fade? We probe only the diffusion module's token representation with linear readouts, so the fade of sequence chemistry there only shows reduced linear accessibility, not that the module loses access to it. We note that the diffusion module is conditioned on the trunk, so it may consume sequence information to compute geometry without ever re-representing it. This is consistent with progressive loss of decodability over diffusion layers and sampling steps (Fig.~\ref{fig:s2_diffgrid} and Fig.~\ref{fig:split}).

The steering experiments (\S\ref{sec:r_steer}) highlight that predictive accessibility and causal influence can come apart. The strand direction is highly predictive yet steers at none of the thirteen layers we tested. The erasure experiments preserved overall secondary-structure composition, but the separate diagnostic shows that labels remain recoverable after analogous erasure. These experiments therefore leave the model's dependence on secondary-structure information unresolved. 

Our hypothesis for the strand-steering null being a result of strand information being carried in the unedited $z$, agrees with \citet{lu_two_2026}, whose steering of pairwise distances in $z$ does induce hairpins in Boltz-1; direct interventions on $z$ are needed to test it.

That coil steers more strongly from the middle of the trunk than from the output that conditions the diffusion module also agrees with \citet{lu_two_2026}, whose pairwise-distance steering in $z$ peaks in middle-to-late blocks. Their charge steering acts in early blocks, but in Boltz-1 these lie in the MSA module, upstream of the Pairformer layers we edit, so our sweep does not test it.

Our comparison of sparse and dense annotations (\S\ref{sec:r_annot}) illustrates the limits of interpretability evaluation for domain-specific ML models. Available annotations of features for biology are often sparse, which makes it difficult to identify encoded features. In a similar vein, we find that held-out supervised probes outperform single SAE features for known concepts and provide more reliable steering directions (\S\ref{sec:r_sae}). We suggest that for interpreting biological ML models, SAEs are most useful for generating hypotheses about unlabeled features, but such claims require independent validation because best-of-many feature selection can inflate scores for rare concepts.

Together, these results motivate three methodological recommendations. First, claims about represented information should specify the readout class and avoid treating linear decodability as direct evidence of causal use. Second, causal conclusions should be restricted to the intervention site and intervention modality that were tested. Third, evaluation against sparse biological annotations should include controls for missing labels whenever possible.

\subsection{Limitations}

This study analyzes one AlphaFold3-class model, one main SAE recipe, a limited set of residue-level labels, and primarily linear readouts. The chemistry contrast rests on SwissProt labels for 99 proteins, so several concepts have wide intervals. Steering targets only secondary structure. Necessity could be tested only at the trunk output, because inside the diffusion module the control edit alone damaged the structure, and the larger-dose check of early layers covers only 16 proteins. Probe folds are grouped by protein but not clustered by sequence identity, though we find train--test sequence-identity leakage negligible (Appendix~\ref{app:methods_sup}). Future work should test whether the trunk/diffusion division holds in other predictors, add extended nonlinear probes, improve SAEs for diffusion activations, map the trunk$\to$diffusion pathway with activation patching, and steer $z$ directly, which would test our explanation of the strand null and extend to pairwise chemistry such as disulfide bridges.

\section*{Responsible-use statement}

This work analyzes representations of an already-open, publicly released model (Boltz-1) and does not increase capability for harm beyond what the base model already provides. Our steering results move secondary-structure content of the model prediction rather than enabling any novel design.

\section*{Acknowledgements and funding}

PJ is supported by the Biotechnology and Biological Sciences Research Council (UKRI-BBSRC) grant BB/T008784/1. OMC acknowledges funding from a New College Todd-Bird Junior Research Fellowship and MRC Fellowship MR/Y010078/1. A. W. is supported by the Wellcome Trust (Grant Number: 218514/Z/19/Z).

% ===========================================================================
% References and appendices are excluded from the 5-page main-text limit
% per the workshop CFP.
% ===========================================================================

\bibliographystyle{plainnat}
\bibliography{references}

\newpage
\appendix
\renewcommand{\thefigure}{S\arabic{figure}}
\setcounter{figure}{0}
\renewcommand{\thetable}{S\arabic{table}}
\setcounter{table}{0}
\makeatletter
\renewcommand{\theHfigure}{S\arabic{figure}}
\renewcommand{\theHtable}{S\arabic{table}}
\makeatother

\section{Appendix}
\label{app:methods_sup}

\subsection{Probe training and scoring}

For each concept, layer, stack, and feature space, we train an independent binary logistic-regression probe. Probes are not multi-task: each concept is fit separately. We use grouped five-fold cross-validation with proteins as groups, so residues from the same protein never appear in both train and test folds. Within each fold, features are standardized using a training-fold StandardScaler. Logistic regression uses fixed inverse regularization $C=1.0$, the lbfgs solver, a maximum of 2000 iterations, and balanced class weights. We do not perform hyperparameter search.

Probe predictions are evaluated out of fold. For comparability with the single-feature readouts, we report the best F1 over a fixed set of percentile thresholds applied to the pooled out-of-fold scores. Thus, the probe metric should be interpreted as a held-out ranking/decodability score rather than as the F1 of a calibrated classifier at threshold 0.5.

Concepts with fewer than five positive residues are skipped. If a training fold contains no positive examples for a concept, that fold is skipped. Rare labels are otherwise handled only through balanced class weights; we do not upsample positives or use stratified protein clustering.

\subsection{F1, precision, recall, and nulls}

All F1 scores use the domain-level metric introduced in prior InterPLM-style evaluations \citep{simon_interplm_2025}. Precision is computed per residue. Recall is computed per domain: a positive domain is counted as recalled if at least one positive prediction overlaps it. This metric rewards detecting the presence of a biological region without requiring exact residue-boundary recovery.

For single-feature readouts, we compute a permutation null by circularly shifting labels and repeating the same best-of-features selection procedure. We define signal as observed F1 minus null F1. Main single-feature comparisons are restricted to informative concepts whose held-out probe-raw F1 clears the corresponding null by more than 0.05 in at least one stack. Because SAE-1feat and neuron-1feat are in-sample best-of-many statistics, they can exceed held-out probes on rare concepts through winner's-curse effects. We therefore use them only as feature-alignment measures.

\subsection{Secondary-structure annotation control}

SwissProt secondary-structure annotations are sparse because they record experimentally curated regions rather than providing a dense label for every residue. To test whether apparent false positives reflect model/probe error or annotation incompleteness, we compare SwissProt secondary-structure labels against dense DSSP labels on the same activations. Dense DSSP labels are computed both from AlphaFold structures and from Boltz-1's own predicted structures. The Boltz-own DSSP labels provide a self-consistency control: if large differences between SwissProt and DSSP arose only because AlphaFold and Boltz-1 disagree structurally, then scoring against Boltz-own DSSP should dramatically change the secondary-structure results.

\subsection{Limitations of the evaluation protocol}

Our train/test folds are grouped by protein but are not clustered by sequence identity. Near-identical homologs could therefore appear in different folds. This may inflate absolute probe F1 values, although the main comparisons use the same split protocol across layers and modules. The readouts measure decodability and feature alignment, not causal use: a concept can be linearly decodable without being necessary for prediction, and a high-scoring SAE latent need not be a faithful causal steering direction. Causal interventions and activation patching are required to test which representations are actively used by the model. Our steering experiments test one such intervention family, additive and ablative edits of $s$, and cannot speak to why sequence chemistry is less linearly accessible in the diffusion module.

\subsection{Sequence-identity leakage control}

To assess whether grouped cross-validation could be inflated by homologous proteins appearing in different folds, we quantified sequence identity between training-fold and held-out test proteins. For each evaluation split, we performed all-vs-all global alignment using BLOSUM62 and recorded train--test protein pairs exceeding 30\% sequence identity. Across the 486 evaluation proteins, only 3.7\% (18 of 486) had any training-fold neighbor above this threshold, and only 9 of approximately 117,000 train--test protein pairs exceeded it. Thus, train--test sequence-identity leakage across folds is negligible and cannot plausibly explain the observed probe-raw performance. Because both modules are evaluated on the same proteins with the same fold assignments, any remaining inflation would affect the Pairformer and diffusion readouts equally and therefore cannot explain the module-specific attenuation of sequence-chemical information.

\subsection{Causal steering protocol}
\label{app:steering}

\begin{figure}[t]\centering
  \includegraphics[width=0.6\textwidth]{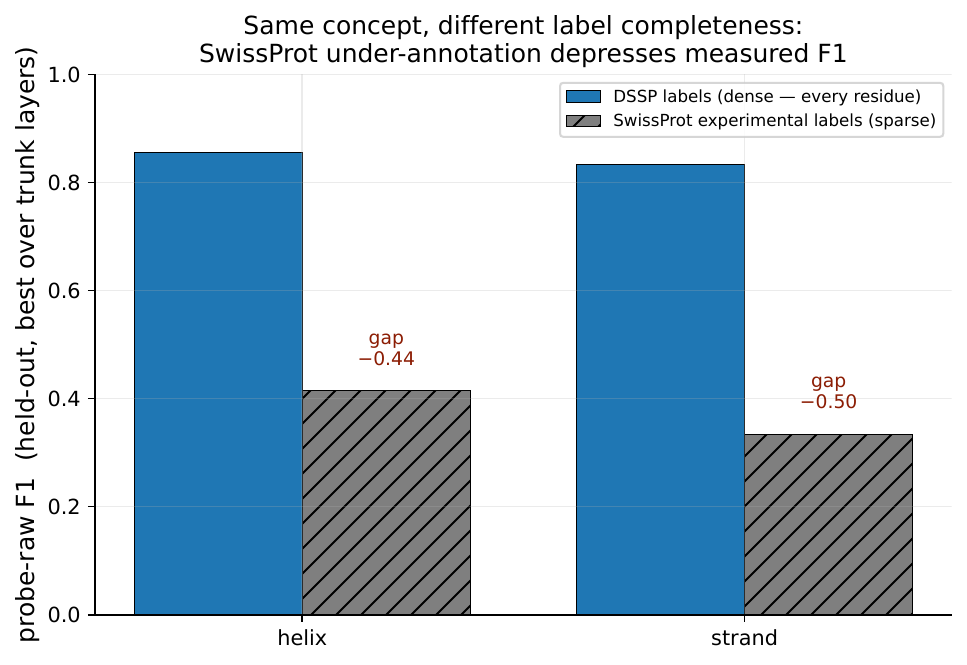}
  \caption{\textbf{F1 against SwissProt is a precision lower bound.} Probe-raw F1 for helix and
  strand on identical trunk activations scored against dense DSSP vs.\ sparse SwissProt
  experimental secondary-structure (SS) annotations (helix $0.85$ vs.\ $0.41$; strand $0.83$ vs.\
  $0.33$; each the best over trunk layers). The gap is SwissProt under-annotation, not decoder failure.}
  \label{fig:underannotation}
\end{figure}

Steering directions are fit only on the 389-protein training split and applied to all 97 held-out proteins over the full dense-DSSP helix gradient (range $0.05$--$0.95$). The train/held-out split is stratified by helix content and seeded, and is fixed before any steering, so the held-out evaluation set is outcome-independent (an earlier 34-protein run used a pre-steering subset of the same 97 and gave the same qualitative pattern with smaller magnitudes, driven by fewer low-pLDDT proteins). Directions are taken at Pairformer layer 47 ($s_\text{trunk}$ conditioning). The probe direction is the held-out logistic-probe weight mapped back to raw-activation space ($\text{coef\_}/\text{scaler.scale\_}$) and $\ell_2$-normalized; the SAE direction is the decoder column of the single highest-F1 latent (helix latent F1 $0.72$, coil $0.71$, strand $0.47$). Additive steering uses $x' = x + k\cdot\text{mean}|(x-\mu)\cdot u|\,u$ with $k\in\{1,2,4,8,16\}$; ablation uses $x' = x - \alpha\,((x-\mu)\cdot u)\,u$. Every condition (baseline, concept-add, random-add, ablation) denoises from identical resampled noise, so effects are not confounded with an independent diffusion sample, and mean pLDDT is monitored as a check that the representations stay within the model's learned distribution of protein representations (stable at $\sim$74 at $k{=}16$: baseline $73.8$, concept-add $73.4$--$73.7$). Effects are reported paired per protein as (concept-add $-$ random-add) on the concept's own DSSP fraction, with 95\% paired-protein bootstrap CIs ($B{=}10{,}000$). To characterize specificity against the high-dimensional space of possible directions, we additionally run a null distribution of 19 matched-norm random directions on the same 97 proteins (a dedicated shard, reused across concepts) and report each concept's population effect as a $z$-score against this null (Table~\ref{tab:steering}, Fig.~\ref{fig:steer_null}). As a vector-validity check, the exact steering directions are re-scored for held-out F1 on the 97 proteins (helix $0.83$, strand $0.78$, coil $0.88$; precision $0.98$, $0.93$ and $0.90$). These are the F1 values the abstract and \S\ref{sec:r_steer} quote, because they score the vector that was actually added on the proteins it was added to; they run a little below the decodability figures reported in \S\ref{sec:r_annot} and \S\ref{sec:r_sae}, which are measured on the probe evaluation set and, in Fig.~\ref{fig:underannotation}, at each concept's best trunk layer. The re-score confirms that the strand null is not due to a failure of the probe to generalize to the steered proteins.

\subsection{Trunk-depth sweep}
\label{app:depth}

\begin{figure}[H]\centering
  \includegraphics[width=\textwidth]{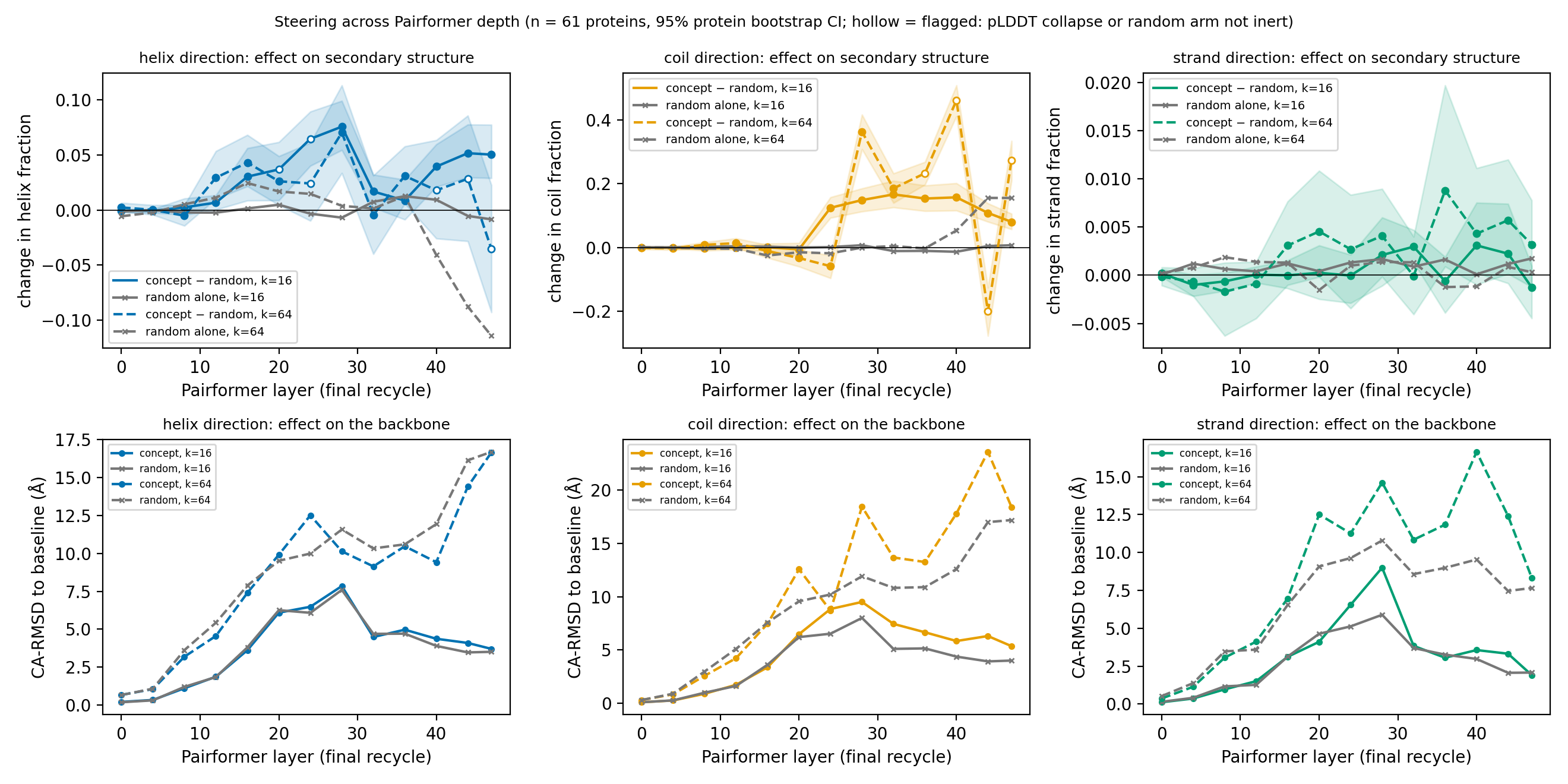}
  \caption{\textbf{Steering across Pairformer depth.} Probe directions added to $s$ at one Pairformer
  layer (final recycle) at $k{=}16$ (solid) and $k{=}64$ (dashed); every legend entry names its dose.
  Each row carries one kind of quantity. \emph{Top, signed composition:} the paired effect, concept
  minus matched-norm random, with its 95\% protein bootstrap CI, together with each arm's own raw
  change. \emph{Middle, reshuffling:} churn, $(|\Delta H|+|\Delta E|+|\Delta C|)/2$, the minimum
  fraction of residues that must have left their baseline DSSP state. Churn is unsigned and so is kept
  off the signed axis above. Both arms are shown: over layers 24--47 at $k{=}16$ the coil direction
  reshuffles $2$--$4$ times as much as its matched control, while the helix direction reshuffles about
  as much as its own --- yet in both cases only the concept arm's signed line moves. What separates a
  concept edit from a random one of the same norm is that its reshuffling has a destination, not that
  there is more of it; a random edit's near-zero signed line is two opposite shifts cancelling, not
  inertness. \emph{Bottom, backbone:} C$\alpha$-RMSD to the baseline structure. Hollow markers in the
  top row flag cells where either arm's mean pLDDT falls by more than 5, or the random direction itself
  moves the state by more than $0.05$ on average. All $n{=}97$ held-out proteins completed every cell.}
  \label{fig:s_depth}
\end{figure}

\begin{figure}[H]\centering
  \includegraphics[width=\textwidth]{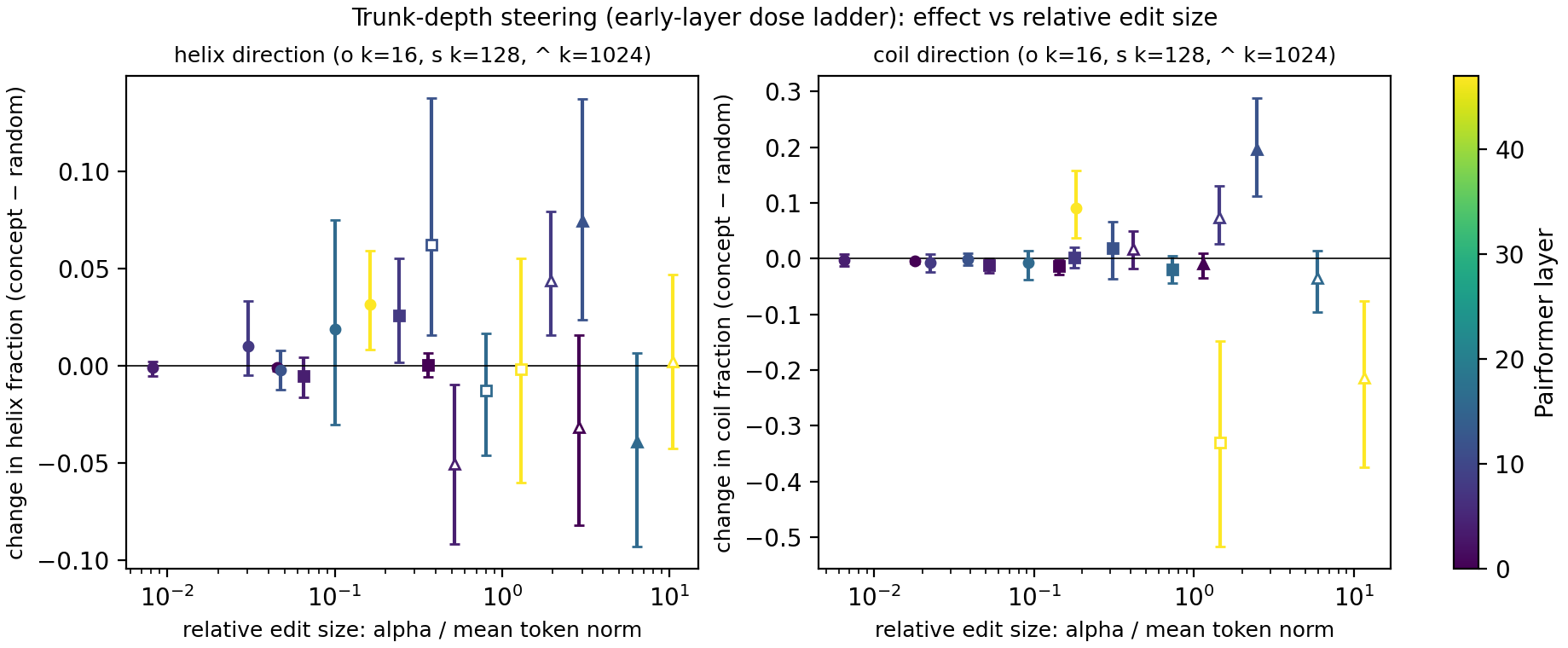}
  \caption{\textbf{Early-layer dose ladder.} The dose $k$ scales with the projection scale at the edit's own layer, which is small relative to $\lVert s\rVert$ early in the trunk, so the fixed $k$ of Fig.~\ref{fig:s_depth} delivers a far smaller relative edit at layers 0--12 (below $0.05$ at $k{=}16$) than at layer 47 ($0.15$--$0.17$), and the early-layer null there could be a dosing artefact. Here layers 0, 4, 8, 12, 16 and 47 are re-run
  for helix and coil at $k\in\{16,128,1024\}$, and effects are plotted against the relative edit size
  $\alpha/\lVert s\rVert$ at the edit's own layer rather than against $k$. Layer 0 shows no
  concept-specific effect at any rung whose arms stay undamaged, including rungs well past the relative
  edit at which layer 47 steers; layer 4's small projection scale caps its undamaged rungs at a
  relative edit of $0.06$, so it is untested rather than null; layer 12 steers once driven and layer 8
  points the same way without surviving correction at $n{=}16$. Hollow markers flag the same damage
  conditions as Fig.~\ref{fig:s_depth}: 8 of the 12 cells at $k{=}1024$, and every layer-47 cell above
  $k{=}16$, where the edit exceeds $\lVert s\rVert$ itself. $n{=}16$ held-out proteins, the shorter and
  more helix-rich head of the list.}
  \label{fig:s_depth_dose}
\end{figure}

For the layer sweep (Fig.~\ref{fig:s_depth}), probe directions are fit in the same way at Pairformer layers 0, 4, \ldots, 44, and the layer-47 direction above is reused, giving 13 layers. Each is added to $s$ at its own layer on the final recycle only, at $k\in\{16, 64\}$ with the projection scale measured at that layer, and paired with a matched-norm random direction. All 97 held-out proteins completed every layer, concept and dose; at layer 47 and $k{=}16$ every protein's DSSP result is identical to the main steering run. A cell is flagged when either arm's mean pLDDT falls by more than 5, or the random arm itself moves the target state by more than $0.05$, since a difference between two damaged predictions is not steering. The additive edit is $k$ times the projection scale, $\text{mean}|(x-\mu)\cdot u|$, measured at the edit's own layer, so a fixed $k$ does not deliver a fixed relative edit $\alpha/\lVert s\rVert$ across depth. Both the projection scale and the mean token norm $\lVert s\rVert$ grow across the trunk, but the projection scale grows faster (about 200-fold for helix and 570-fold for coil, against 62-fold for $\lVert s\rVert$). At $k{=}16$ the relative edit is therefore below $0.05$ at layers 0--12, against $0.15$--$0.24$ from layer 20 onwards, and the main text compares layers at a relative edit similar to that of the output's steering edit ($0.15$--$0.19$). We additionally re-ran layers 0, 4, 8, 12, 16 and 47 for helix and coil at $k\in\{16,128,1024\}$ on 16 of the held-out proteins, the first 16 in run order, which are shorter and more helix-rich than the full set (mean baseline helix fraction $0.52$ against $0.46$; Fig.~\ref{fig:s_depth_dose}).

\subsection{Necessity protocol}
\label{app:necessity}

We test two ablations of the single representation $s$ at the last Pairformer layer, which conditions the diffusion module. First, a rank-one ablation removes a single probe direction and is compared with a matched-norm random ablation on the 97 held-out proteins. Second, a self-fit erasure equalises the helix, strand and coil class means within each protein, using DSSP labels from that protein's unedited predicted structure. Its control applies an edit of the same overall size using labels circularly shifted along the sequence by half the protein's length. The erasure run covers 95 of the 97 held-out proteins; two were excluded because DSSP label coverage was below $0.95$.

\begin{figure}[H]\centering
  \includegraphics[width=\textwidth]{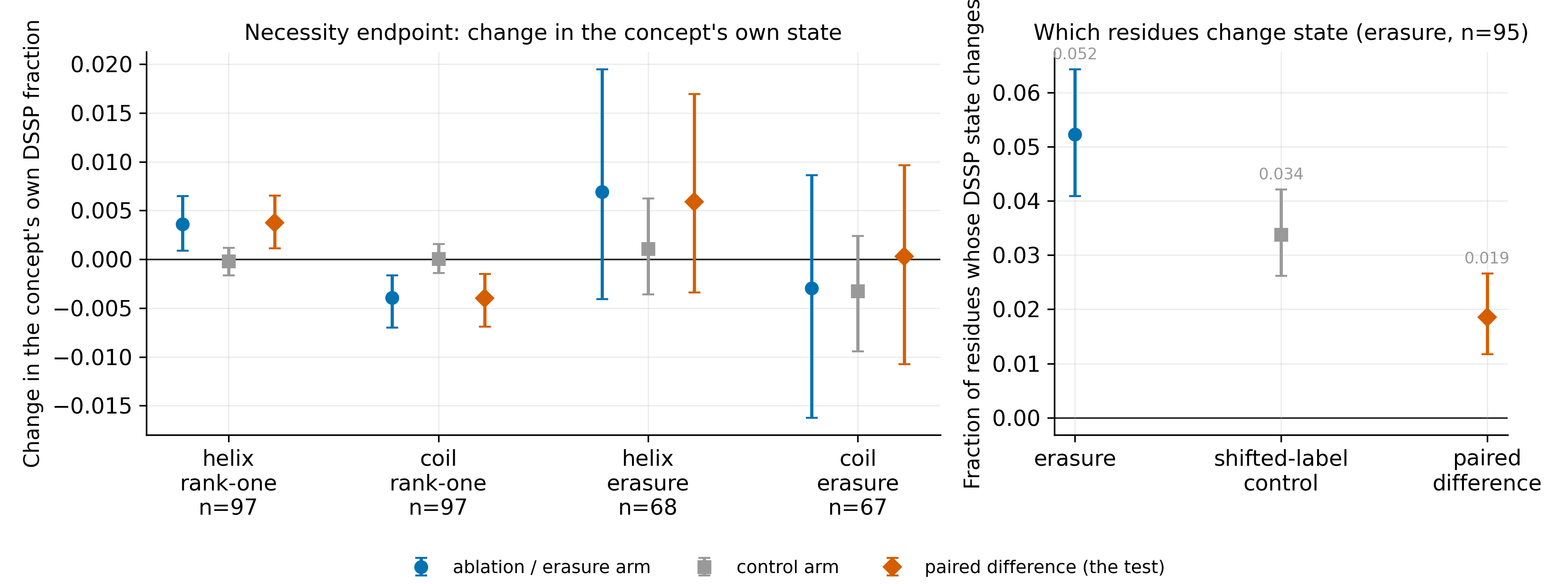}
  \caption{\textbf{Necessity is null at the trunk conditioning.} Two ablations of $s_\text{trunk}$, the
  tensor the diffusion module is conditioned on, differing about fourfold in how far they move the
  structure (median C$\alpha$-RMSD $0.4$ against $1.6$~\AA). Blue: the ablation or erasure arm's change
  from the unedited baseline; grey: its control arm; vermillion: the paired per-protein difference,
  which is the test. 95\% protein bootstrap CIs. \emph{Left:} the concept's own DSSP fraction. Neither intervention substantially reduces overall helix or coil content. The rank-one ablation moves the fraction by at most $0.004$ ---
  raising helix and, for coil, lowering it by $-0.004$ [$-0.007$, $-0.001$], two orders of magnitude
  short of removal --- against the $+0.039$ and $+0.061$ the same directions produce when added at this
  site; the erasure's intervals exclude a loss larger than about one point. \emph{Right:} the erasure is
  not inert at the residue level, moving $5.2\%$ of residues out of their baseline state against $3.4\%$
  for its control. The run's pre-registered positive control, a label swap, does move the endpoint
  ($-0.040$ [$-0.061$, $-0.019$] helix) but is not interpretable, because its own shifted-label control
  breaks the damage floor (median C$\alpha$-RMSD $6.2$~\AA). The diffusion-module sites were run on the same proteins but
  are excluded here: there the shifted-label control alone moves structures by a median
  $7.9$--$10.8$~\AA{}, past the $2$~\AA{} damage floor fixed in advance, so no necessity test is
  possible.}
  \label{fig:s_necessity}
\end{figure}

The analysis plan specified that a site is interpretable only if its control keeps median C$\alpha$-RMSD within $2$~\AA{} and median pLDDT drop within 5 points. The trunk control satisfies these criteria, with median C$\alpha$-RMSD $1.3$~\AA{}. At the diffusion-module sites, control medians are $7.9$--$10.8$~\AA{}, so those interventions do not support a necessity conclusion and are excluded from Fig.~\ref{fig:s_necessity}. For the trunk erasure, the helix and coil confidence intervals have half-widths of approximately $0.0100$ and $0.0103$, respectively, compared with the planned half-width of $0.010$. The exclusion of losses larger than one percentage point is therefore approximate.

We additionally assessed decodability after analogous erasure in a separate protein-centred diagnostic on 95 proteins, using independently extracted activations and DSSP labels. Residues belonging to classes too small to be edited were excluded. We report within-protein AUROC, pooling positive--negative residue comparisons only within proteins to avoid effects from between-protein composition. Refitted linear logistic probes returned constant predictions (AUROC $0.500$), as expected from equalised class means.

Using the same quadratic readout pipeline based on 48 principal components before and after erasure, AUROC decreased from $0.940$ to $0.652$ for helix, $0.956$ to $0.754$ for strand, and $0.898$ to $0.673$ for coil. A control repeating the procedure with labels permuted within proteins, including refitting the erasure, yielded AUROC $0.493$--$0.499$. Thus, erasure reduced quadratic decodability without eliminating above-chance prediction. This diagnostic does not replay the exact structural intervention or establish whether Boltz-1 uses the information accessible to these readouts. Preservation of overall helix and coil content therefore leaves the necessity of secondary-structure information unresolved.

\subsection{Probe bootstrapping}
\label{app:probe bootstrap}

To quantify uncertainty for sparse SwissProt concepts, we use a cluster bootstrap over proteins, treating proteins as the independent unit. For each concept, we first compute the full-sample operating point exactly as in the main benchmark, then fix that point during resampling. We resample proteins with replacement 2000 times, recompute per-residue precision, per-domain recall, and F1 from per-protein sufficient statistics, and report percentile 95\% confidence intervals. We do not re-select thresholds or features inside each bootstrap sample, because that would compound selection noise with sampling uncertainty. This bootstrap is used for both trunk and diffusion evaluations.

\section{Supplementary figures}

\label{app:suppfigs}

\begin{figure}[H]\centering
  \includegraphics[width=0.7\textwidth]{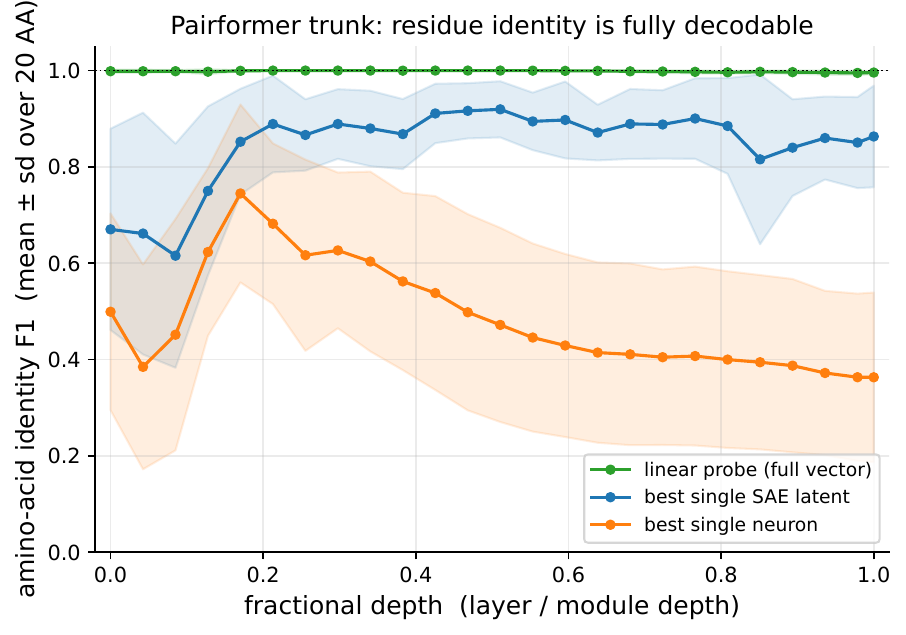}
  \caption{\textbf{Calibration on amino-acid identity (positive control).}
  Held-out probe-raw recovers residue identity at $\text{F1}\approx1.0$ across
  all trunk depths ($0.996$ at output), confirming label/activation alignment.
  A single SAE latent captures identity cleanly (SAE-1feat $\approx0.86$;
  Cys $\approx0.90$ at the output, peaking at $\approx0.99$ in the mid-trunk)
  while the best single neuron lags ($\approx0.36$). This
  fixes the F1 reference scale for all later panels.}
  \label{fig:calibration}
\end{figure}

\begin{figure}[H]\centering
  \includegraphics[width=\textwidth]{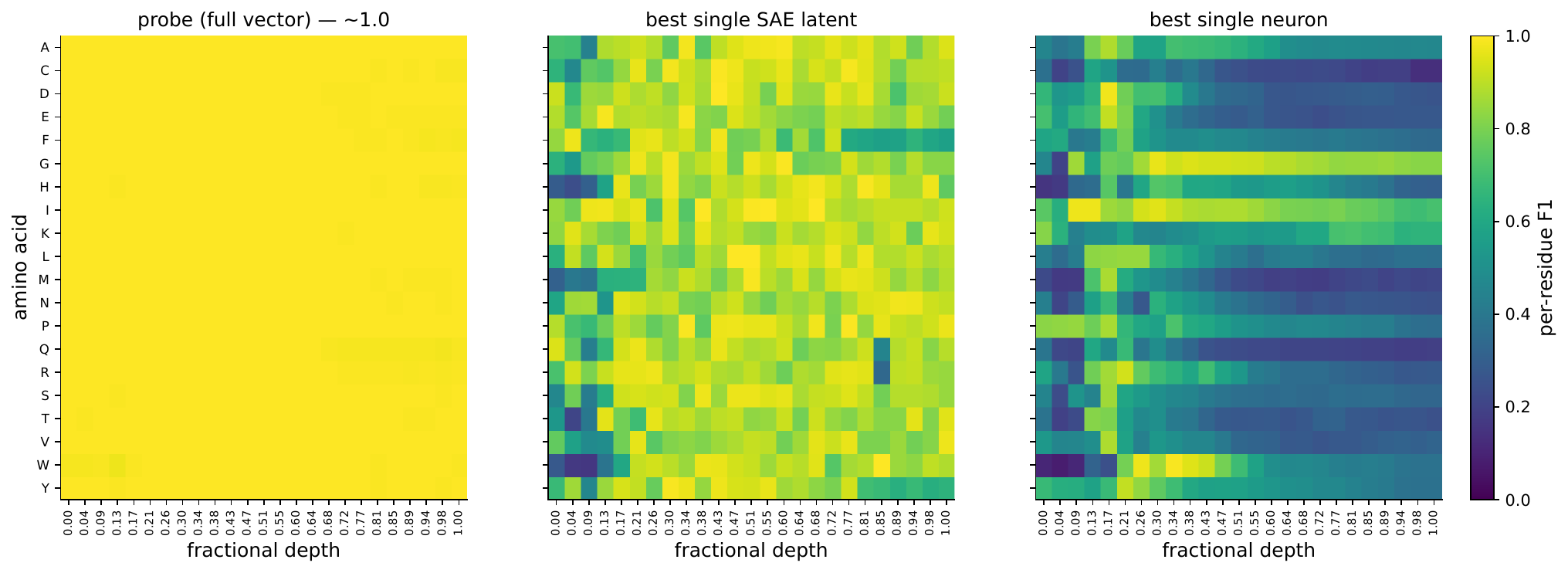}
  \caption{\textbf{Pairformer per-amino-acid identity breakdown.} (Left) held-out probe-raw
  F1 for each of the 20 residues across trunk depth, confirming alignment is
  correct per residue, not merely on average. (Right) SAE-1feat vs.\
  neuron-1feat per residue; the SAE concentrates identity into dedicated latents
  (e.g.\ Cys $\approx0.99$ at its mid-trunk peak) the raw basis does not expose.}
  \label{fig:s1_peraa}
\end{figure}

\begin{figure}[H]\centering
  \includegraphics[width=\textwidth]{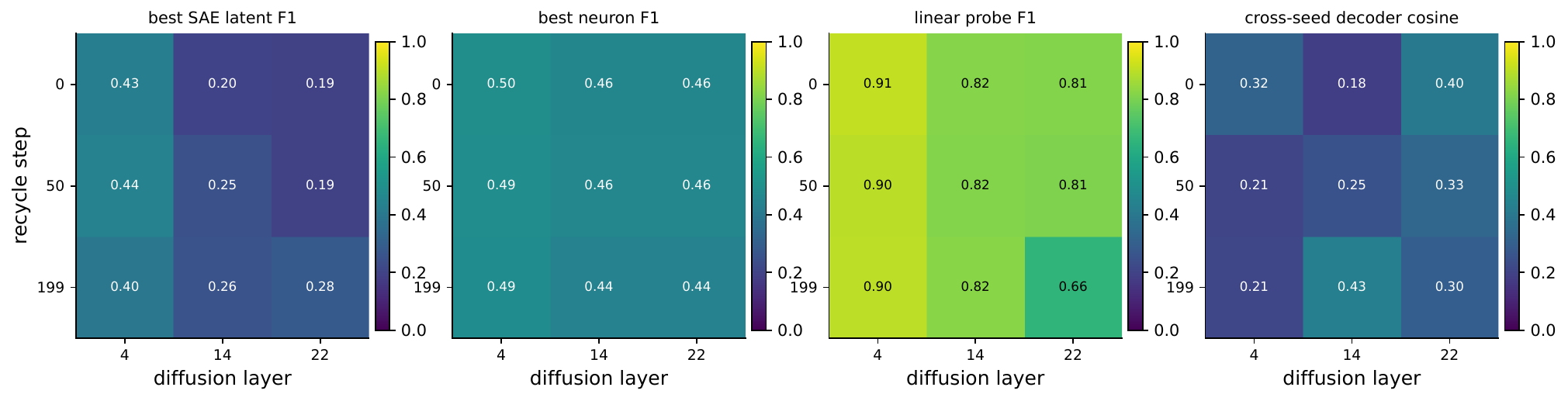}
  \caption{\textbf{Diffusion module per-amino-acid identity breakdown.} (Left) held-out probe-raw
  F1 for each of the 20 residues across diffusion module depth.}
  \label{fig:s9_peraa}
\end{figure}

\begin{figure}[H]\centering
  \includegraphics[width=\textwidth]{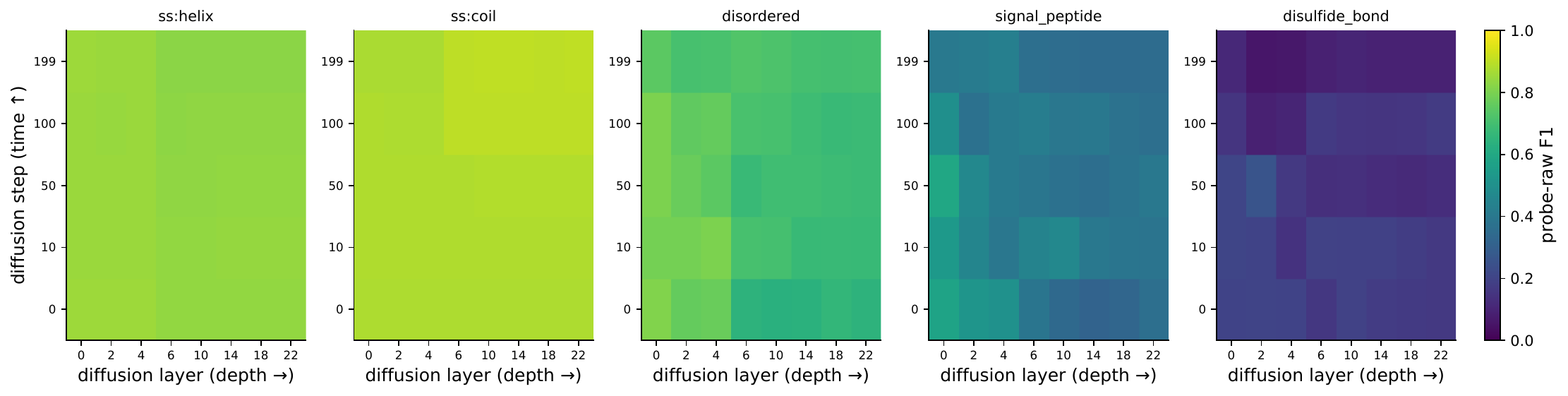}
  \caption{\textbf{Full diffusion layer$\times$sampling-step heatmaps}
  (extends Fig.~\ref{fig:split}B,C). Geometry concepts are bright across the
  grid; sequence-chemistry concepts are dark, with a gradual fade along both
  axes for signal peptide.}
  \label{fig:s2_diffgrid}
\end{figure}

\begin{figure}[t]\centering
  \includegraphics[width=0.7\textwidth]{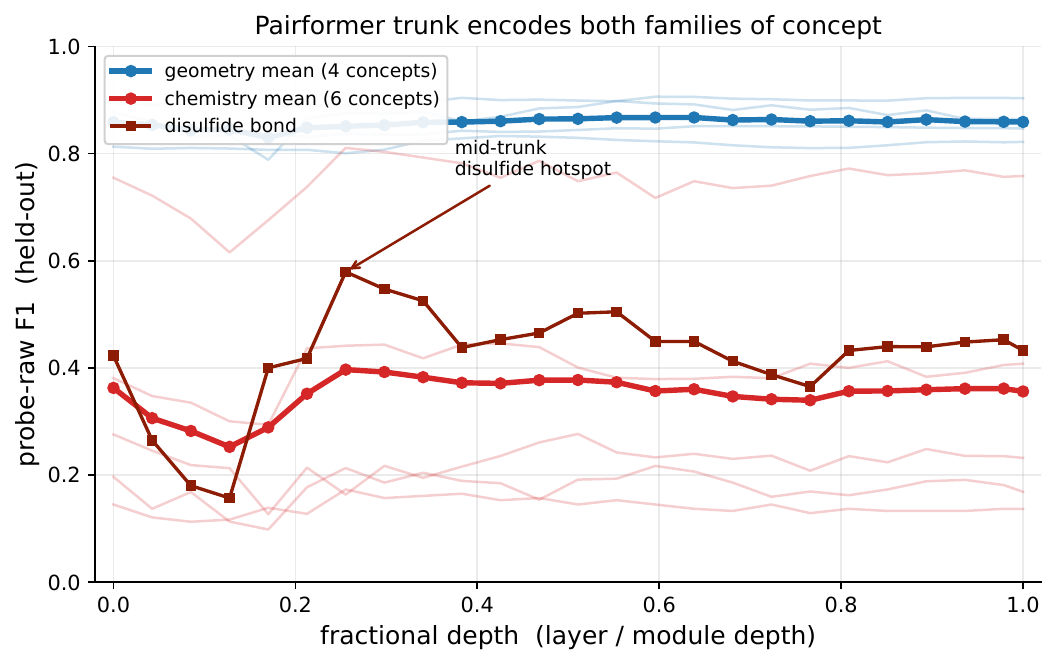}
  \caption{\textbf{The Pairformer trunk encodes geometry and chemistry.}
  Held-out probe-raw F1 vs.\ fractional trunk depth (recycle~1) for secondary
  structure (helix/strand/coil, $\approx0.79$--$0.90$) and sequence-chemical
  concepts (signal peptide $\approx0.76$, disorder $\approx0.86$, disulfide
  $\approx0.43$ with a mid-trunk hotspot at depth $\approx0.2$--$0.35$). The
  trunk is not geometry-only.}
  \label{fig:trunk}
\end{figure}

\begin{figure}[H]\centering
  \includegraphics[width=\textwidth]{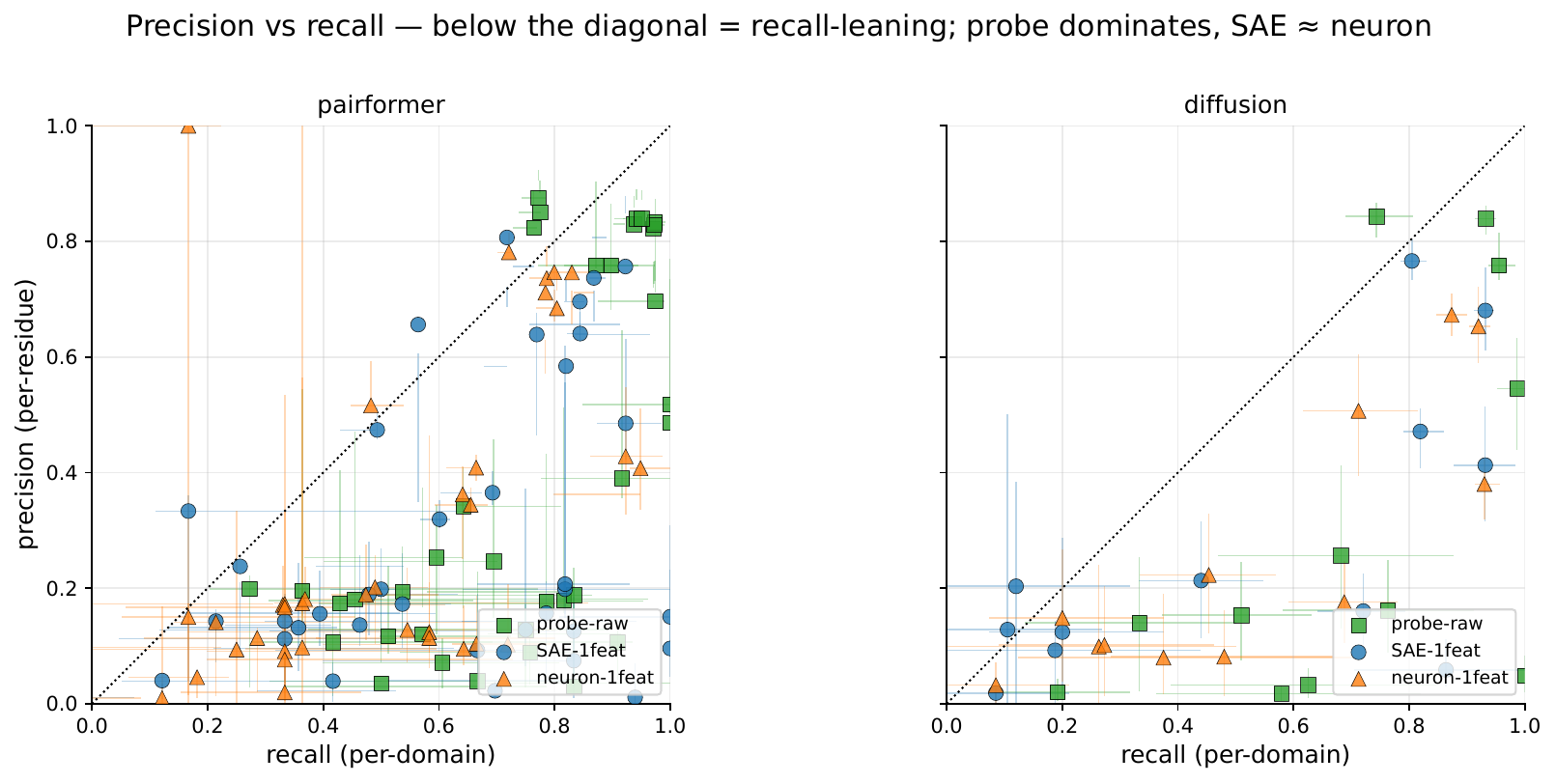}
  \caption{\textbf{Probes dominate; single features are recall-leaning.}
  Precision vs.\ recall for the three readouts with both measured, in both stacks. The held-out probe
  dominates both axes (trunk P0.40/R0.74; diffusion P0.32/R0.69); most points
  fall below the precision\,$=$\,recall diagonal (broad, noisy detectors), the
  exception being probe-raw on DSSP strand, which sits above it at every trunk
  layer and in diffusion. The
  best single SAE latent beats the best neuron by only $\approx+0.03$ F1 (trunk)
  / $\approx-0.01$ (diffusion).}
  \label{fig:pr}
\end{figure}

\begin{figure}[H]\centering
  \includegraphics[width=\textwidth]{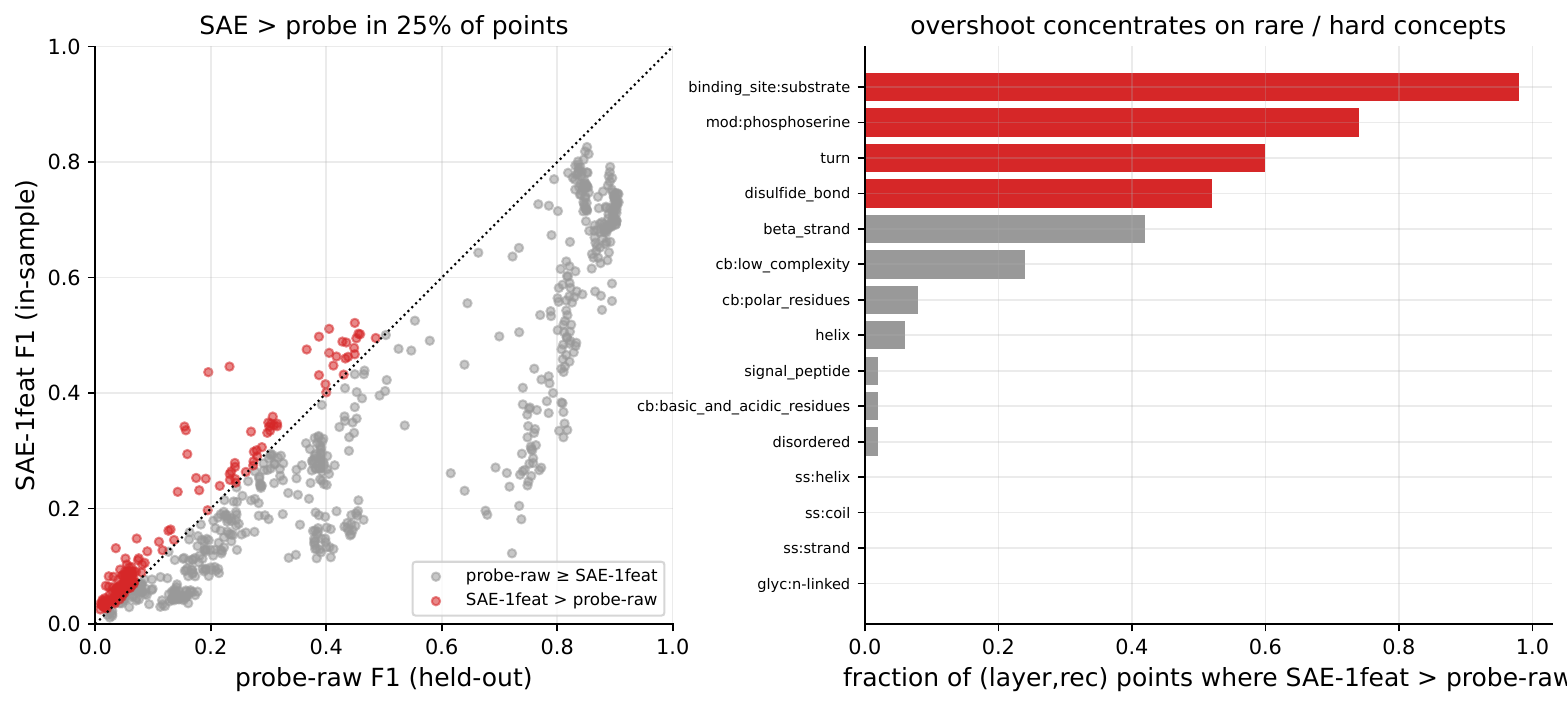}
  \caption{\textbf{Winner's-curse diagnosis.} Cases where SAE-1feat exceeds the held-out probe concentrate on rare concepts where the probe is correctly near zero; $\text{signal}=\text{F1}-\text{null}$ collapses for these, so SAE-1feat indexes monosemanticity, not decodability. Genuine high-F1 features (e.g.\ disulfide) are exempt.}
  \label{fig:s4_winnerscurse}
\end{figure}

\begin{figure}[H]\centering
  \includegraphics[width=\textwidth]{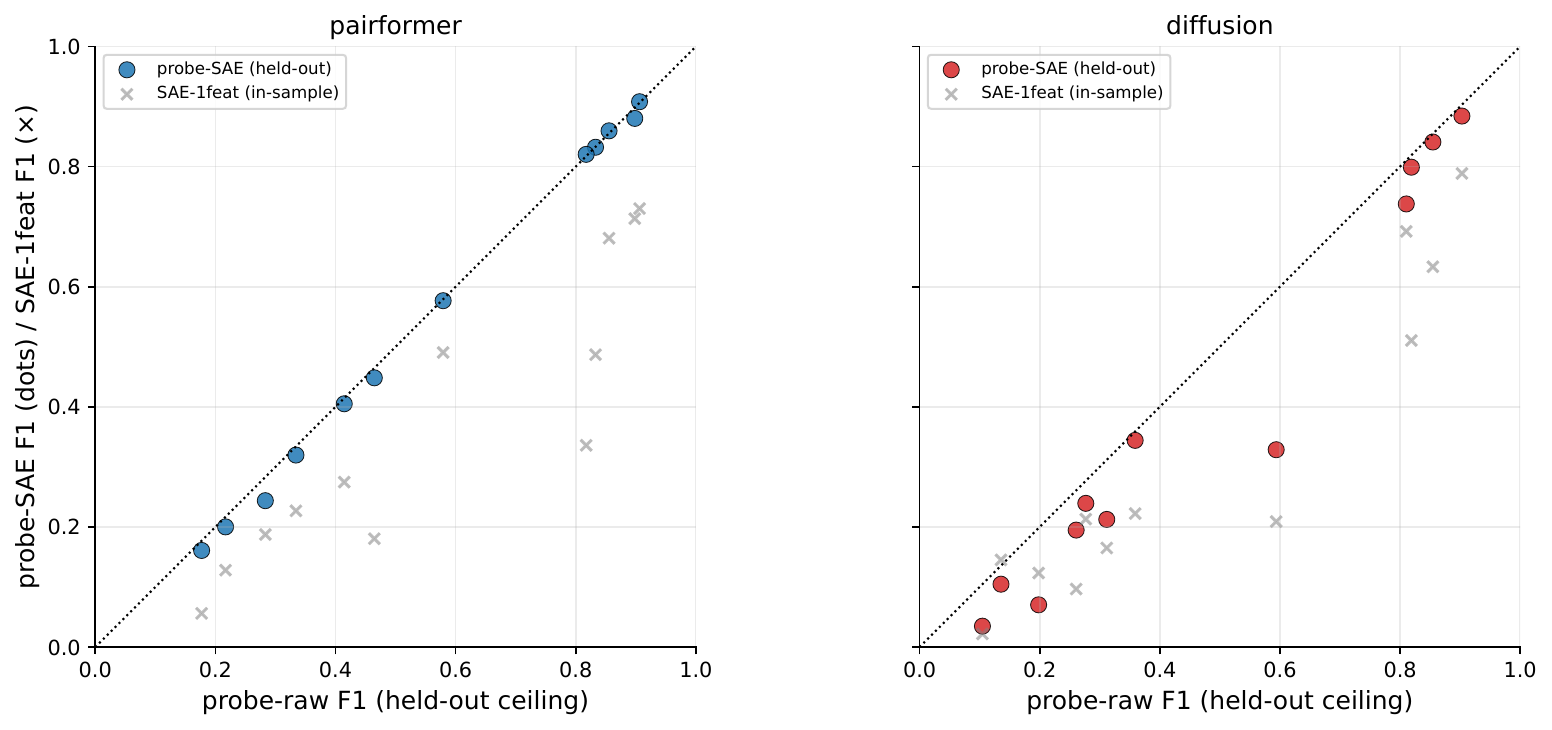}
  \caption{\textbf{SAE faithfulness} (probe-SAE\,/\,probe-raw) vs.\ depth.
  $\sim$0.97 in the trunk but only $\sim$0.82 in diffusion, so single-feature F1
  understates the diffusion module specifically.}
  \label{fig:s5_faithfulness}
\end{figure}

\begin{figure}[H]\centering
  \includegraphics[width=\textwidth]{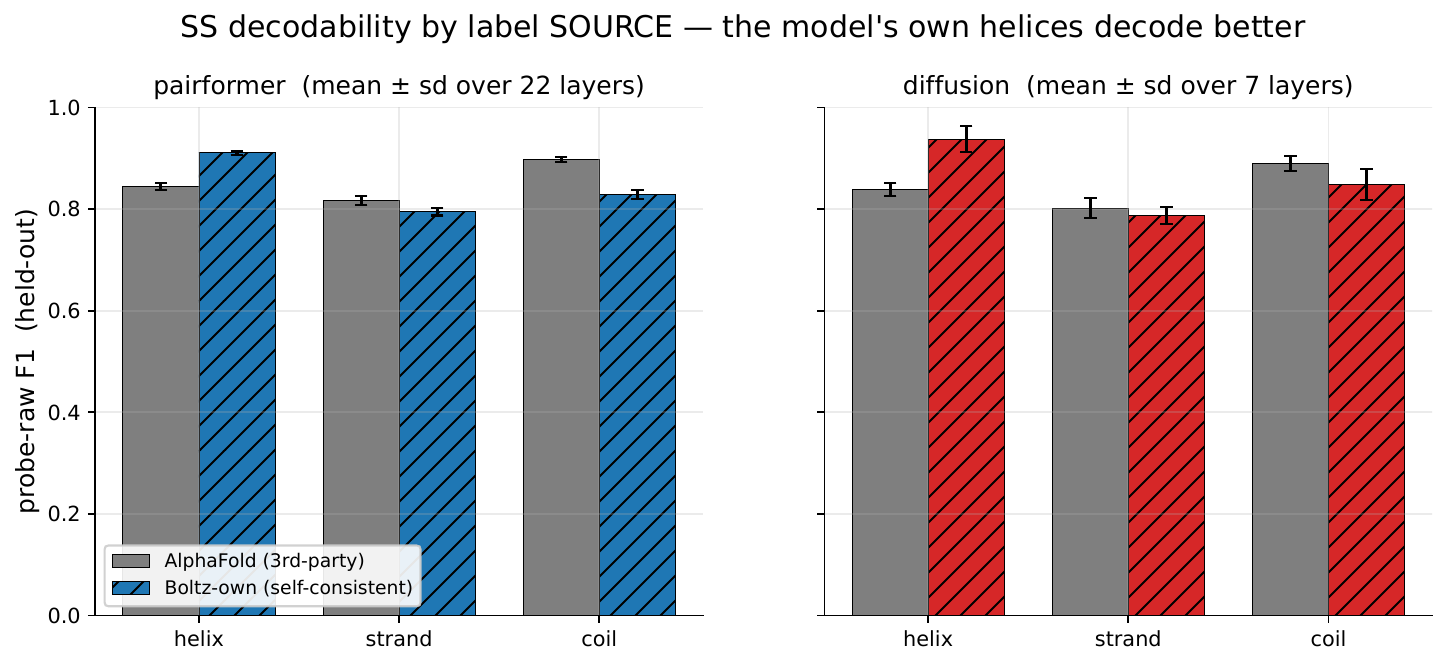}
  \caption{\textbf{Self-consistency control.} Scoring against DSSP on Boltz's
  own predicted structure raises helix probe-raw by only $\sim$0.06--0.08 over
  AlphaFold-derived labels at every depth in both stacks (AlphaFold slightly
  better on strand/coil), so the large SwissProt deltas reflect under-annotation,
  not model disagreement.}
  \label{fig:s7_selfconsistency}
\end{figure}

\begin{figure}[H]\centering
  \includegraphics[width=\textwidth]{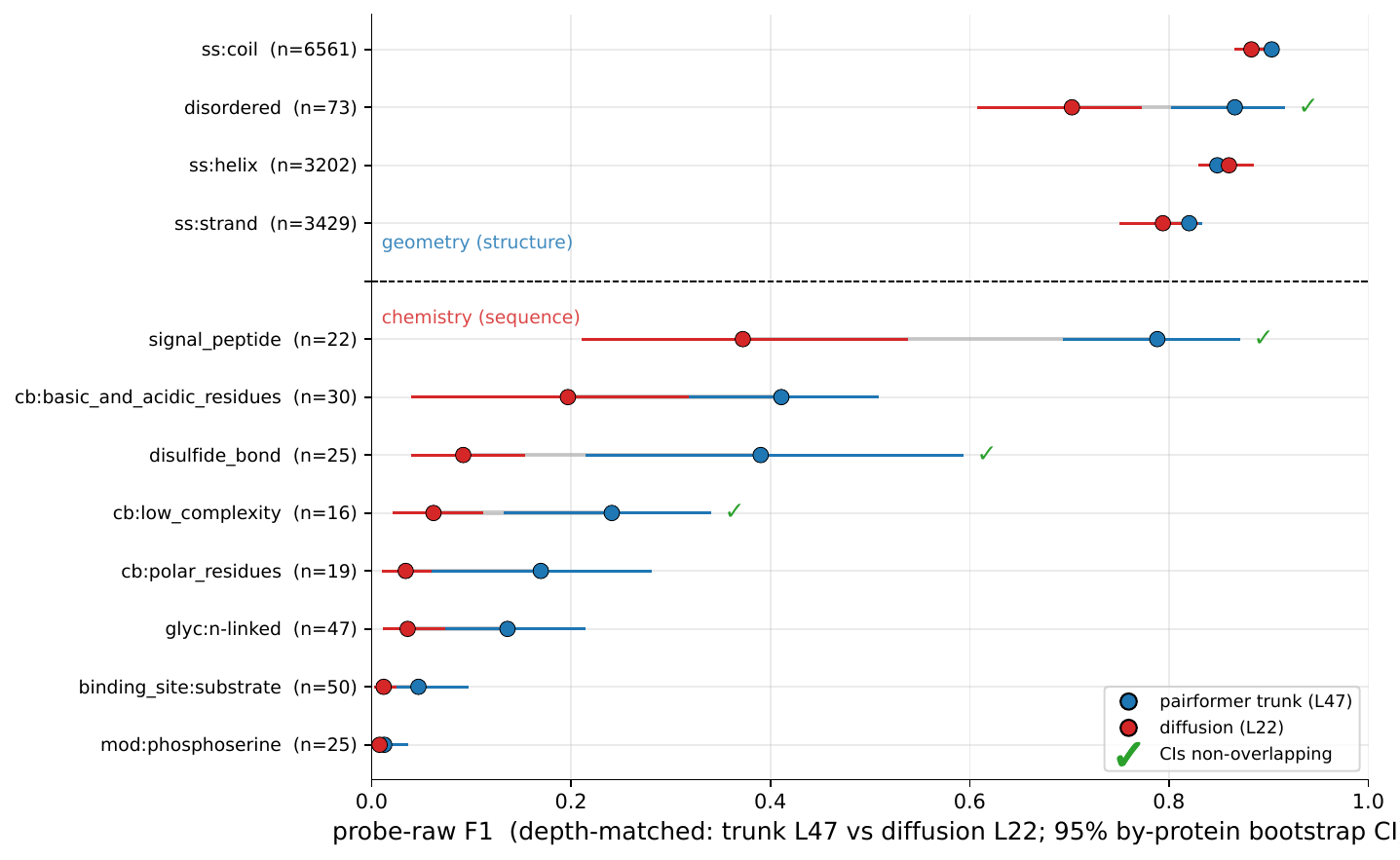}
  \caption{\textbf{Depth-matched probe-raw F1 with 95\% protein-bootstrap confidence intervals.}   Pairformer trunk (L47) and diffusion module (L22) outputs are compared at matched depth using the same fixed operating point, with 95\% cluster-bootstrap CIs over proteins. Geometry concepts (ss:helix, ss:strand, ss:coil) remain similar across modules, whereas sequence-chemistry concepts drop in the diffusion module. The rare SwissProt labels have wider intervals, but the chemistry-vs-geometry split remains clear.}
  \label{fig:s11_probe_CI}
\end{figure}

% ---- causal-steering supplementary figures ---------------------------------
\begin{figure}[H]\centering
  \includegraphics[width=\textwidth]{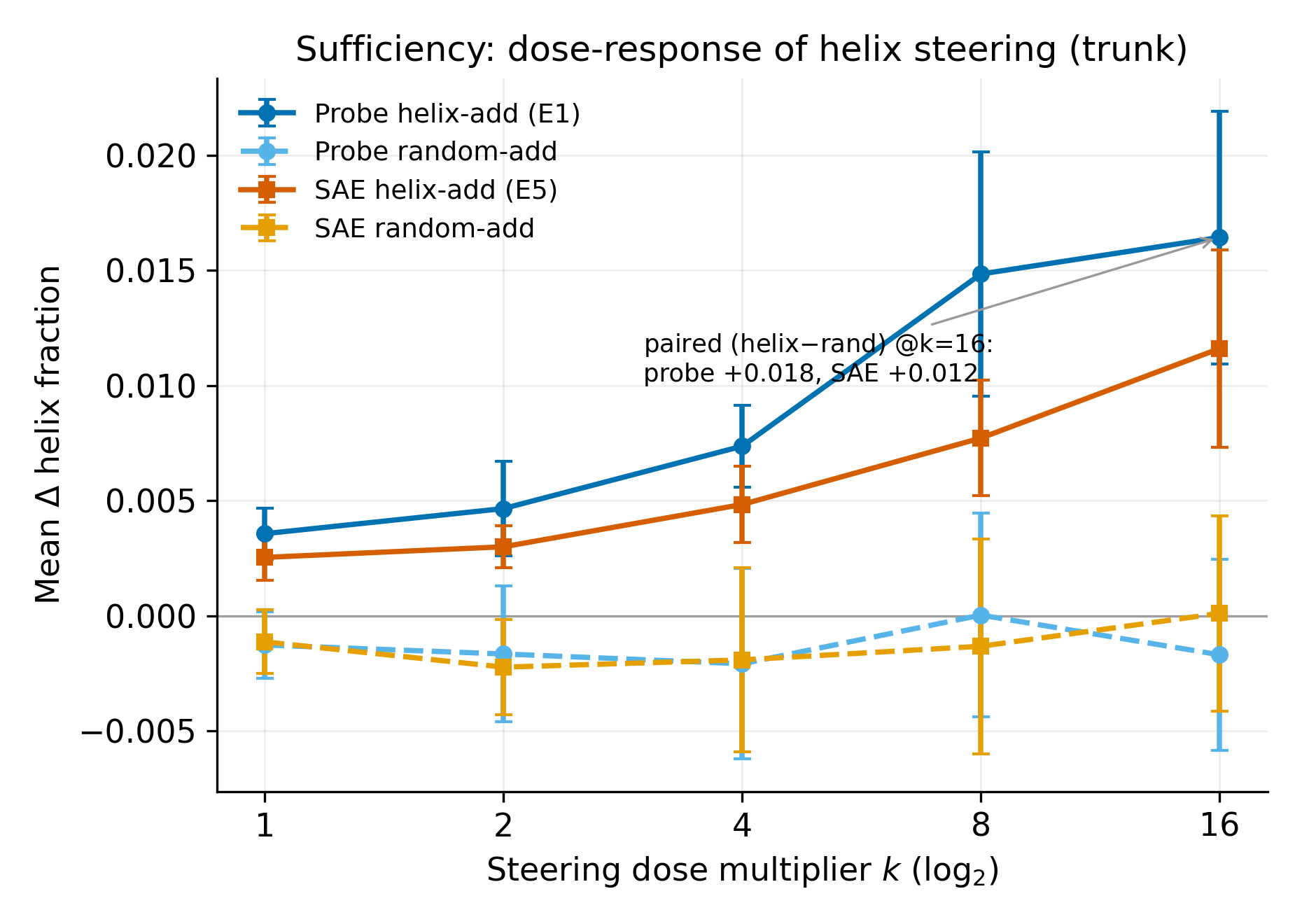}
  \caption{\textbf{Steering is dose-dependent for helix and coil, not strand.} \emph{Left:} the paired
  per-protein effect on each concept's own DSSP fraction, concept-add minus its matched-norm random
  control, against additive dose $k\in\{1,2,4,8,16\}$, for the supervised probe direction (solid) and
  the best single SAE latent (dashed), with 95\% paired-protein bootstrap CIs. The helix and coil probe
  directions grow monotonically with dose; the strand probe and the coil and strand latents stay at zero
  throughout, so the strand null is not a dosing artefact. \emph{Right:} mean pLDDT of the same arms,
  drawn on a scale that spans the $5$-point drop treated as damage elsewhere in this paper; every arm
  stays within one point of the unedited baseline, so the response is graded and within the model's
  learned distribution of protein representations. $n{=}97$ held-out proteins; the $k{=}16$ column
  reproduces Table~\ref{tab:steering} exactly.}
  \label{fig:s_dose}
\end{figure}

\begin{figure}[H]\centering
  \includegraphics[width=0.7\textwidth]{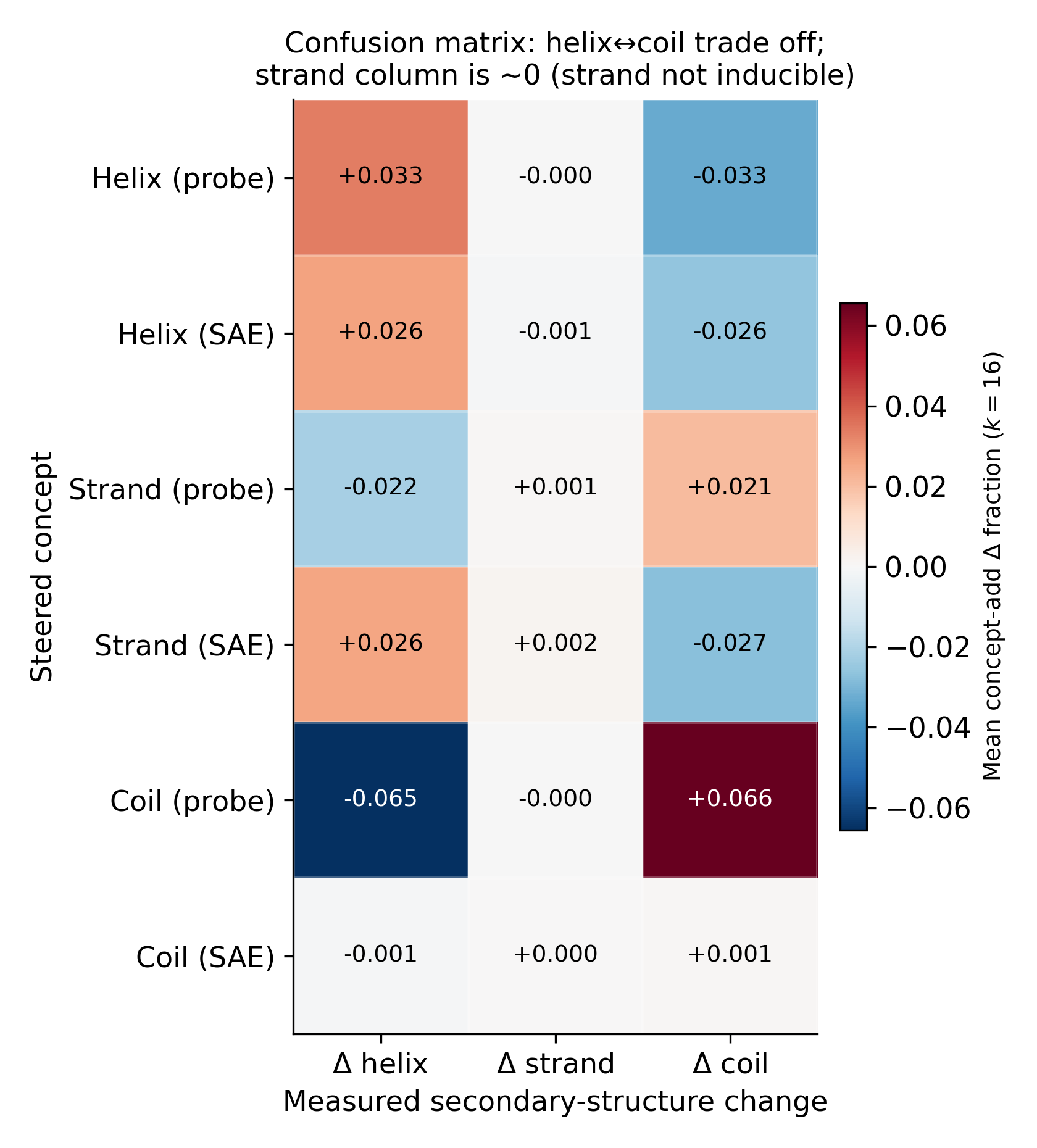}
  \caption{\textbf{Steering is specific: a helix$\leftrightarrow$coil
  anti-diagonal.} Mean concept-add effect (dose $k{=}16$) of each steered
  direction (rows) on each measured DSSP state (columns). Helix and coil occupy
  the anti-diagonal---steering one raises it and lowers the other---while the
  strand column is $\approx0$ throughout (strand is neither induced nor a
  byproduct).}
  \label{fig:steer_confusion}
\end{figure}

\begin{figure}[H]\centering
  \includegraphics[width=\textwidth]{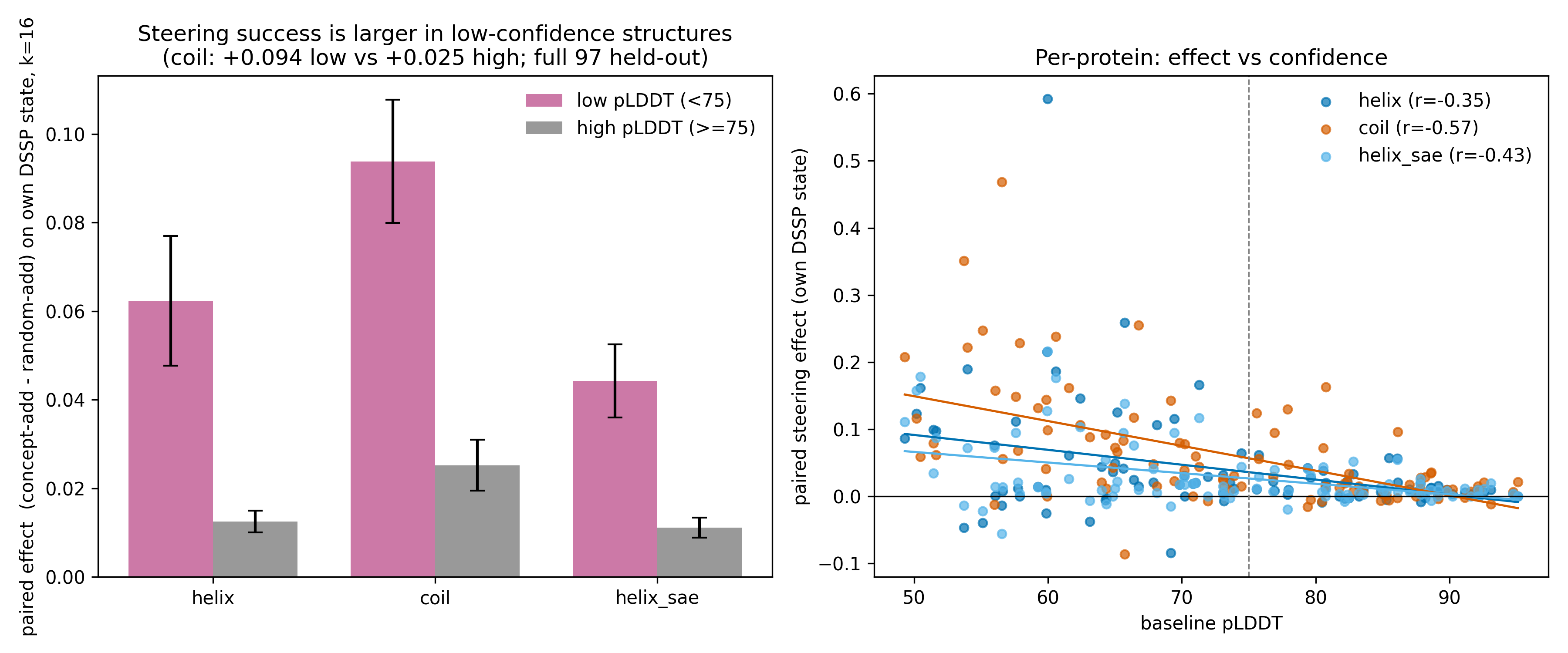}
  \caption{\textbf{Steering success scales with model uncertainty.} Paired effect
  by baseline pLDDT bin (low $<75$, $n{=}51$ vs.\ high $\ge75$, $n{=}46$).
  Low-confidence structures steer far more (coil $+0.094$ vs.\ $+0.025$; helix
  $+0.062$ vs.\ $+0.013$); the per-protein effect correlates with pLDDT at
  $r\approx-0.57$ for coil. The conditioning matters most where the fold is not
  yet committed.}
  \label{fig:steer_confidence}
\end{figure}

\section{Supplementary tables}

\begin{table}[H]\centering
  \caption{Depth-matched module outputs (probe-raw F1 unless noted). Geometry
  transfers; sequence chemistry is reduced. pf~L47 is the Pairformer trunk output,
  diff~L22 the deepest extracted diffusion token-transformer layer.}
  \label{tab:headline}
  \small
  \begin{tabular}{lccc}
    \toprule
    Concept & Trunk (pf L47) & Diffusion (L22) & $\Delta$ \\
    \midrule
    Coil           & 0.90 & 0.90 & $\,$0.00 \\
    Helix          & 0.85 & 0.83 & $-0.02$ \\
    Strand         & 0.82 & 0.82 & $\,$0.00 \\
    Disordered     & 0.86 & 0.70 & $-0.16$ \\
    Signal peptide & 0.76 & 0.35 & $-0.40$ \\
    Disulfide bond & 0.43 & 0.10 & $-0.34$ \\
    AA identity (probe)     & 1.00 & 0.66 & $-0.34$ \\
    AA identity (SAE-1feat) & 0.86 & 0.30 & $-0.57$ \\
    \bottomrule
  \end{tabular}
\end{table}

\begin{table}[H]\centering
  \caption{Paired steering effects on the concept's own DSSP state (held-out,
  dose $k{=}16$, $n{=}97$ proteins). Effect $=$ per-protein
  (concept-add $-$ random-add) with 95\% paired-protein bootstrap CI
  ($B{=}10{,}000$); $z$ is the population effect against a 19-direction
  matched-norm random null (Fig.~\ref{fig:steer_null}). Mean pLDDT stays stable
  ($\sim$74) across conditions.}
  \label{tab:steering}
  \small
  \begin{tabular}{llccc}
    \toprule
    Concept & Direction & Paired effect (95\% CI) & $z$ vs.\ null & Verdict \\
    \midrule
    Coil   & probe     & $+0.061\ [+0.045,+0.079]$ & $+11.5$ & steers (strongest) \\
    Helix  & probe     & $+0.039\ [+0.024,+0.056]$ & $+6.2$  & steers \\
    Helix  & SAE-1feat & $+0.029\ [+0.020,+0.038]$ & $+5.0$  & steers \\
    Strand & probe     & $-0.001\ [-0.003,+0.001]$ & $+0.3$  & null \\
    Strand & SAE-1feat & $+0.001\ [-0.001,+0.003]$ & $+1.0$  & null \\
    Coil   & SAE-1feat & $+0.001\ [-0.005,+0.007]$ & $-0.1$  & null \\
    \bottomrule
  \end{tabular}
\end{table}

\section{Dataset and label prevalence}
\label{app:dataset_label_prevalence}

The labeled evaluation sets differ by annotation source. SwissProt functional annotations are sparse and are available for fewer proteins than dense DSSP secondary-structure labels. We therefore report both dataset sizes and per-concept label prevalence. Positive domains are maximal contiguous positive residue runs and are the unit used for domain-level recall in all F1 calculations.

\begin{table}[ht]
\centering
\small
\caption{Dataset size and per-concept label prevalence for the evaluation sets.}
\label{tab:dataset_label_prevalence}
\resizebox{\textwidth}{!}{%
\begin{tabular}{llrrrrr}
\toprule
\textbf{Dataset} & \textbf{Concept} & \textbf{Positive Residues} & \textbf{Positive Domains} & \textbf{Proteins w/ Concept} & \textbf{Total Proteins} & \textbf{Total Residues} \\
\midrule
SwissProt & Beta Strand & 2,033 & 352 & 23 & 99 & 44,968 \\
          & Binding Site: Substrate & 83 & 50 & 15 & 99 & 44,968 \\
          & Compositional Bias: Basic \& Acidic & 506 & 30 & 18 & 99 & 44,968 \\
          & Compositional Bias: Low Complexity & 278 & 16 & 13 & 99 & 44,968 \\
          & Compositional Bias: Polar Residues & 279 & 19 & 12 & 99 & 44,968 \\
          & Disulfide Bond & 1,179 & 25 & 12 & 99 & 44,968 \\
          & Glycosylation: N-Linked GlcNAc & 47 & 47 & 13 & 99 & 44,968 \\
          & Helix & 2,810 & 295 & 23 & 99 & 44,968 \\
          & Modified Residue: Phosphoserine & 26 & 25 & 10 & 99 & 44,968 \\
          & Region: Disordered & 3,949 & 73 & 39 & 99 & 44,968 \\
          & Signal Peptide & 480 & 22 & 22 & 99 & 44,968 \\
          & Turn & 275 & 80 & 22 & 99 & 44,968 \\
\midrule
DSSP Secondary & Secondary Structure: Helix & 56,977 & 4,760 & 393 & 393 & 154,299 \\
               & Secondary Structure: Strand & 18,449 & 5,108 & 324 & 393 & 154,299 \\
               & Secondary Structure: Coil & 78,873 & 9,787 & 393 & 393 & 154,299 \\
\midrule
Boltz Secondary & Secondary Structure: Helix & 98,574 & 7,737 & 485 & 486 & 196,136 \\
                & Secondary Structure: Strand & 24,706 & 6,982 & 428 & 486 & 196,136 \\
                & Secondary Structure: Coil & 72,856 & 14,526 & 486 & 486 & 196,136 \\
\bottomrule
\end{tabular}%
}
\end{table}

\end{document}